\documentclass[aps,prb,reprint,amsmath,amssymb,groupedaddress,onecolumn]{revtex4-1}

\usepackage{longtable}
\usepackage[pdftex]{graphicx}
\usepackage{bm}
\usepackage{amsmath}

\begin{document}


\title{Enhanced micropolar model for wave propagation in ordered granular materials}


\author{A. Merkel}
\email{aurelien.merkel.etu@univ-lemans.fr}
\author{S. Luding}
\affiliation{Multi Scale Mechanics, TS, CTW, UTwente, P.O. Box 217, 7500 AE Enschede, Netherlands}
\email{aurelien.merkel.etu@univ-lemans.fr}


\date{\today}

\begin{abstract}
The vibrational properties of a face-centered cubic granular crystal of monodisperse particles are predicted using a discrete model as well as two micropolar models, first the classical Cosserat and second an enhanced Cosserat-type model, that properly takes into account all degrees of freedom
at the contacts between the particles.
The continuum models are derived from the discrete model via a micro-macro transition of the discrete relative displacements and particle rotations to the respective continuum field variables.
Next, only the long wavelength approximations of the models are compared and, considering the discrete model as reference, the Cosserat model shows inconsistent predictions of the bulk wave dispersion relations. This can be explained by an insufficient modeling of sliding mode of particle interactions in the Cosserat model. An enhanced micropolar model is proposed including only one new elastic tensor from the more complete second order gradient micropolar theory. This enhanced micropolar model then involves the minimum number of elastic constants to consistently predict the dispersion relations in the long wavelength limit. 
\end{abstract}


\maketitle

\section{Introduction}
The classical theory of elasticity consists of a macroscopic material description. The material is not described at the micro-level by considering the displacement of the different particles in interaction, but is described as a continuum by considering macroscopic quantities as stress and strain. The classical elasticity theory can be viewed as a first gradient of the displacement field approximation of solid state theory \cite{ashcroft1976} and is thus valid in the long wavelength limit only. 
Granular media, due to their micro-inhomogeneous character, are not well described by the standard continuum theory of elasticity. 
In contrast to classical continua, where the sizes of the vibrating atoms can be assumed to be negligible compared to the macroscale, the sizes of the particles in a granular assembly are comparable to it \cite{schwartz1984}.
 In addition, considering the sliding, twisting and rolling resistances at the level of the contacts between the particles, a consistent description of the elasticity of a granular medium needs to take into account all the rotational degrees of freedom of each individual particle and thus all the relative degrees of each pair. \\
Considering this, the elastic behaviors of crystalline structures of monodisperse beads can be efficiently described by a discrete model, where the displacement and rotation of each individual bead are taken into account. One of the major differences with classical elasticity is the existence of optical-type rotational-related modes of wave propagation. Especially, the dispersion relations given by the discrete model of the bulk waves propagating in crystalline structures of contacting monodisperse beads, without solid bridges between them, is consistent with experimental results \cite{merkel2010PRE,merkel2011prl} in a hexagonal close-packed structure, and with numeric simulation results in a face-centered cubic structure \cite{mouraille2006,mouraille2008efm,mouraille2008phd}. 
Nevertheless, the discrete model can be solved analytically only for well-know regular crystalline structures, the case of a random assembly of beads can be done \cite{kruyt2010,kruyt2012} but is too complex for large systems. Moreover, due to diffraction scattering and attenuation effects, even for a small level of randomness, only long wavelength waves will propagate in a granular assembly \cite{mouraille2008,dazel2010}. More specifically, only the long wavelength waves can propagate in as a coherent ballistic wave \cite{jia1999,jia2004,langlois2014}. 
Considering that only the long wavelength waves will propagate in random assemblies, which differs from the ideal crystalline case, and that their discrete modeling is too complex, a continuum formulation seems more suitable. A continuum model is also relevant when considering the wave propagation in a granular assembly in contact with an elastic solid \cite{wallen2015}. \\
The generalization of the classical elasticity theory accounting for the rotational degrees of freedom of bodies is known as the Cosserat theory \cite{cosserat1909,eringen1999}. But, from a direct comparison between the dispersion relations predicted with a discrete model and with the ones of a Cosserat model, the latter does not predict correctly the dispersion relations of the modes of propagation related to the rotational motion of the beads \cite{merkel2011prl}. Similarly, the simulation results of a Couette shear cell differ from the predictions of a Cosserat model \cite{mohan1999,latzel2001,mohan2002,latzel2003phd}. Despite many theoretical efforts, the comparison between experimental results and predictions from the Cosserat theory remains inconclusive; a consistent continuum description of granular assemblies is still challenging, see \cite{maugin2010,goddard2014} and the references therein.  
As it will be shown below, one of the interactions, or relative degrees of freedom, between the beads involving the rotational degrees of freedom is not modeled properly leading to inconsistent results. The drawback of the Cosserat theory can be overcome using a second-order gradient theory \cite{muhlhaus1996,suiker2001-1,suiker2001-2,suiker2005}. Nevertheless, the second-order gradient micropolar theory introduces three new elastic tensors, which involve too many new elastic constants to represent a feasible alternative to the discrete modeling. \\
In this work, after an identification of the drawbacks of the Cosserat model, we propose a model based on a continuum formulation that correctly describes the wave propagation in granular media in the long wavelength limit. 
In Sec. \ref{secdiscrete} and in order to get a reference for the comparison of the continuum models, the general theoretical evaluation of the bulk wave propagation in a granular assembly using a discrete model, which follows the derivation in \cite{merkel2010PRE}, is presented. The different interactions between the beads due to contact forces and torques are discussed. 
In Sec. \ref{secdispdiscrete}, the dispersion relations of the bulk eigenmodes propagating in a Face-Centered Cubic (FCC) structure along the $x$-axis are derived using the discrete model. The long wavelength approximations of the dispersion relations, which can be directly compared with the predictions of the continuum models, are then derived. Two cases of contacts between the beads are considered. In the first one, the contacts are considered without solid bridges between the beads and the surface roughness is negligible; this case is called the \textit{frictional case} corresponding to normal and sliding resistant contacts. In the second one, the contacts between the beads are considered with solid bridges and this case is called the \textit{rolling and twisting resistant case}. 
In Sec. \ref{secsimu}, the dispersion relations of the discrete model are compared to those obtained trough numerical simulations of wave propagation in a FCC structure. 
In Sec. \ref{seccontinuum}, the macroscopic continuum models are derived from the microscopic relations of the discrete model following the homogenization techniques proposed in \cite{suiker2001-1,suiker2001-2}.
In Sec. \ref{secCosserat}, the dispersion relations of the bulk eigenmodes in a FCC structure are derived with the Cosserat model. The problems and drawbacks of this model are then discussed. 
Finally in Sec. \ref{secEnhanced}, the enhanced micropolar model is presented and it is shown that its approximations for small wavenumber are exactly equal to those of the discrete model in both frictional, rolling and twisting resistant cases. 


\section{Description of the problem, starting from the discrete theory}
\label{secdiscrete}

\begin{figure}[ht!]
\centering
\includegraphics[width=13cm]{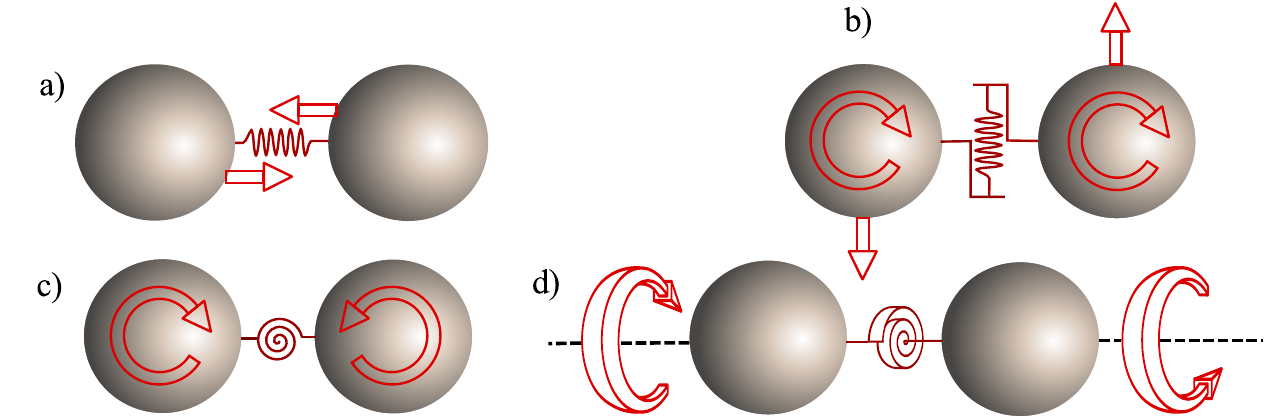}
\caption{(Color online) Sketch of the interactions (relative degrees of freedom) between two beads: a) due to normal resistance, b) due to sliding resistance, c) due to rolling resistance and d) due to twisting (or torsion) resistance. }
\label{figinteraction}
\end{figure}

An assembly of monodisperse beads is considered, all of them being composed by the same material. The diameter of the beads is $a$, the mass density of the material constituting the beads is $\rho_b$, its Poisson's ratio is $\nu$. The mass of one bead with homogeneous density is $m_b=\pi \rho_b a^3/6$, its moment of inertia is $I_b=m_ba^2/10$. The problem is considered in Cartesian coordinates with unit vectors $(\hat{\textbf{x}}, \hat{\textbf{y}}, \hat{\textbf{z}})$. The position of a bead $\alpha$ is defined by the vector $\textbf{R}^{\alpha}$. A local coordinate system $(\textbf{n},\textbf{s},\textbf{t})$ at the level of the surface of contact between two beads is defined: The unit vector $\textbf{n}$, normal to the surface of contact between two beads $\alpha$ and $\beta$, is defined as \cite{merkel2010PRE, chang1995,chang1995IJSS}
\begin{equation}
\textbf{n}=(\textbf{R}^{\beta}-\textbf{R}^{\alpha})/|\textbf{R}^{\beta}-\textbf{R}^{\alpha}|=\cos\phi \hat{\textbf{x}}+ \sin\phi\cos\theta \hat{\textbf{y}}+\sin\phi\sin\theta\hat{\textbf{z}} \simeq (\textbf{R}^{\beta}-\textbf{R}^{\alpha})/a,  
\end{equation}
where it is assumed that the static and dynamic overlaps between the particles are negligible compared to their diameter, $\phi=\arccos (\textbf{n}\cdot\hat{\textbf{x}})$, $\theta=\arccos(\textbf{n}\cdot\hat{\textbf{y}}/\sin\phi)$ if $\phi\neq0$ and $\theta=0$ if $\phi=0$ \cite{chang1995}. 
The two unit vectors $\textbf{s}$ and $\textbf{t}$, which are in the contact plane, are defined as
\begin{eqnarray}
\textbf{s} &= &\partial\textbf{n}/\partial{\phi}= -\sin\phi \hat{\textbf{x}}+ \cos\phi\cos\theta \hat{\textbf{y}}+\cos\phi\sin\theta\hat{\textbf{z}}, \: \nonumber \\
\textbf{t}&=&\textbf{n} \times \textbf{s}=-\sin\theta \hat{\textbf{y}}+\cos\theta\hat{\textbf{z}}. 
\end{eqnarray}
The infinitesimal displacement of bead $\alpha$ is $\textbf{u}^{\alpha}$, its infinitesimal angular rotation is $\textbf{w}^{\alpha}$. 
The dispersion relations are deduced below from the equations of motion for translation
\begin{equation}
m_b\frac{\partial^2 \textbf{u}^{\alpha}}{\partial t^2}=\sum_{\beta} \textbf{F}^{\beta\alpha}, 
\label{eqmottrans}
\end{equation}
and rotation
\begin{equation}
I_b\frac{\partial^2 \textbf{w}^{\alpha}}{\partial t^2}=\sum_{\beta} \textbf{M}^{\beta\alpha} + \frac{1}{2} \sum_{\beta} \textbf{D}^{\beta\alpha}\times\textbf{F}^{\beta\alpha}, 
\label{eqmotrot}
\end{equation}
where the summation is over all the beads $\beta$ in contact with the bead $\alpha$ and the branch vector is
\begin{equation}
\textbf{D}^{\beta\alpha}= \textbf{R}^{\beta}-\textbf{R}^{\alpha}. 
\label{eqDab}
\end{equation}
The direct use of the branch vector in Eq. (\ref{eqDab}) in the equation of motion for rotation in Eq. (\ref{eqmotrot}) is only valid in the case of monodisperse beads, i.e., $|\textbf{R}^{\beta}|=|\textbf{R}^{\alpha}|$, see \cite{suiker2001-1,chang1995,chang1995IJSS,luding2008} for more general formulations. From the linearization of the Hertz-Mindlin contact model between two beads, the contact interactions can be modeled by using a normal and a shear stiffness $K_N$ and $K_S$, respectively \cite{duffy1956,Bjohnson1985,thornton1991,gilles2003}. It should be noticed that this excludes all the nonlinear effects from the analysis \cite{nesterenko2001,tournat2010AAA}. In the case where all the monodisperse beads are composed by the same material, the ratio of shear to normal stiffness is given by $\Delta_K=K_S/K_N=2(1-\nu)/(2-\nu)$. Since the overlap of the beads in contact is small, also the diameter of the surface of contact $d$ is assumed to be small compared the diameter of the beads, e.g. $d\ll a$. Considering the projections on the $x,y,z$ axis of the force applied by bead $\beta$ on bead $\alpha$ as \cite{merkel2010PRE,suiker2001-1}
\begin{eqnarray}
F_i^{\beta\alpha} & = & K_Nn_in_j(u_j^{\beta}-u_j^{\alpha})+K_S(s_is_j+t_it_j)\biggl[u_j^{\beta}-u_j^{\alpha}  + \frac{1}{2}\varepsilon_{jkl}D_k^{\beta\alpha}(w_l^{\beta}+w_l^{\alpha})\biggl] \nonumber \\
& = & \biggl[K_Nn_in_j+K_S(s_is_j+t_it_j)\biggl]\biggl[u_j^{\beta}-u_j^{\alpha} +\frac{1}{2}\varepsilon_{jkl}D_k^{\beta\alpha}(w_l^{\beta}+w_l^{\alpha})\biggl] \nonumber \\
&  = & K^{\beta\alpha}_{ij} \delta^{\beta\alpha}_j,  
\label{eqforce}
\end{eqnarray}
where $K^{\beta\alpha}_{ij}=K_Nn_in_j+K_S(s_is_j+t_it_j)$ and $\delta^{\beta\alpha}_j=u_j^{\beta}-u_j^{\alpha} +\varepsilon_{jkl}D_k^{\beta\alpha}(w_l^{\beta}+w_l^{\alpha})/2$ are the stiffness matrix and the relative displacement, respectively, and $\varepsilon_{ijk}$ is the permutation symbol. \\
While the unit of the siffness constants $K_N$ and $K_S$ is the Newton per meter (N/m), the bending stiffness $G_B$ and the torsion stiffness $G_T$ of the contact are considered here as linear stiffness constants and have the unit Newton meter per radians (Nm/rad) and they include a length squared, i.e., $G_T=K_Ta^2$ and $G_B=K_Ba^2$. A contact model including rolling (or bending) and twisting (or torsion) resistances in agreement with experimental results is beyond the scope of this paper, see \cite{luding2008,brendel2011,fuchs2014} for more details. From the equivalent bending stiffness $G_B$ and torsion stiffness $G_T$ of the contact, the torque applied by bead $\beta$ on bead $\alpha$ is 
\begin{eqnarray}
M_i^{\beta\alpha} & = & G_Tn_in_j(w_j^{\beta}-w_j^{\alpha})+G_B(s_is_j+t_it_j)(w_j^{\beta}-w_j^{\alpha}), \nonumber \\ 
& = &  \bigl[G_Tn_in_j+G_B(s_is_j+t_it_j)\bigl]\bigl(w_j^{\beta}-w_j^{\alpha}\bigl) = G^{\beta\alpha}_{ij}\theta_{j}^{\beta\alpha},  
\label{eqmoment}
\end{eqnarray}
where $G^{\beta\alpha}_{ij}=G_Tn_in_j+G_B(s_is_j+t_it_j)$ and $\theta_{j}^{\beta\alpha}=w_j^{\beta}-w_j^{\alpha}$ are the rotational stiffness matrix and the relative rotation, respectively. 
From Eqs. (\ref{eqforce}) and (\ref{eqmoment}), it is possible to sketch the degrees of freedom involved in the different interactions as displayed in Fig. \ref{figinteraction}. The terms depending on $u_j^{\beta}-u_j^{\alpha}$ are the classical translational normal and shear interactions, which come from the normal and sliding resistance, respectively.
The terms depending on $w_j^{\beta}-w_j^{\alpha}$ are the moment interactions, which are related to the torsion (or twisting) resistance parallel to the vector $\textbf{n}$ and the rolling resistance in the plane composed by the vectors $\textbf{s}$ and $\textbf{t}$. The term depending on $w_l^{\beta}+w_l^{\alpha}$ is the rotational contribution to the translational surface shear displacement, which activates the sliding resistance. The bead rotation is here transformed into a relative displacement in the plane of the surfaces of the spheres through a vectorial cross product. Considering a frictional assembly without solid bridges between the beads, with a negligible roughness of the contact surface, and due to the small size of the contact surface between the beads, the equivalent bending and twisting stiffness are assumed to be negligible, e.g. $G_b=G_t\simeq0$, leading to $M_i^{\beta\alpha}\simeq0$. The bending and the twisting stiffness become non negligible in a rolling and twisting resistant assembly. 


\section{Dispersion relations in the FCC structure in the $x=x_1$ direction}
\label{secdispdiscrete}

This section is dedicated to the discrete model and its long wavelength limit. \\
The FCC structure is a vertical packing of hexagonal contacting layers A, B and C, which are in the closest position related to each other in an ABCABC... sequence. It can also be seen as a vertical packing of cubic contacting layers A and B, which are in the closest position related to each other, in an ABAB... sequence \cite{ashcroft1976}. Without loss of generality, the $x=x_1$ axis is chosen to be perpendicular to the cubic layers, the $y=x_2$ and $z=x_3$ axes are in the plane and along the cubic layers. The position of the beads can be found using integer indices $\alpha_1$, $\alpha_2$ and $\alpha_3$ with 
\begin{equation}
\textbf{R}^{\alpha_1,\alpha_2,\alpha_3}=a[\alpha_1\hat{\textbf{x}}/\sqrt{2}+(\alpha_2+\alpha_1/2)\hat{\textbf{y}}+(\alpha_3+\alpha_1/2)\hat{\textbf{z}}]. 
\label{eqvecposition}
\end{equation}
The FCC structure is transversely isotropic in the plane perpendicular to the $x$ direction. With Eqs. (\ref{eqmottrans})-(\ref{eqmoment}) and after a plane wave substitution $\textbf{u}=\textbf{U}e^{i(\omega t-\textbf{k}\cdot\textbf{x})}$ and $\textbf{w}=\textbf{W}e^{i(\omega t-\textbf{k}\cdot\textbf{x})}$, the dynamical matrix is built. Its eigenvalues give the dispersion relations of the bulk modes that propagates within the crystal with their corresponding eigenvectors $\textbf{X}=(U_x,U_y,U_z,W_x,W_y,W_z)$. 


\subsection{Frictional case}
\label{subsecdispdisc}

\begin{figure}[ht!]
\centering
\includegraphics[width=8cm]{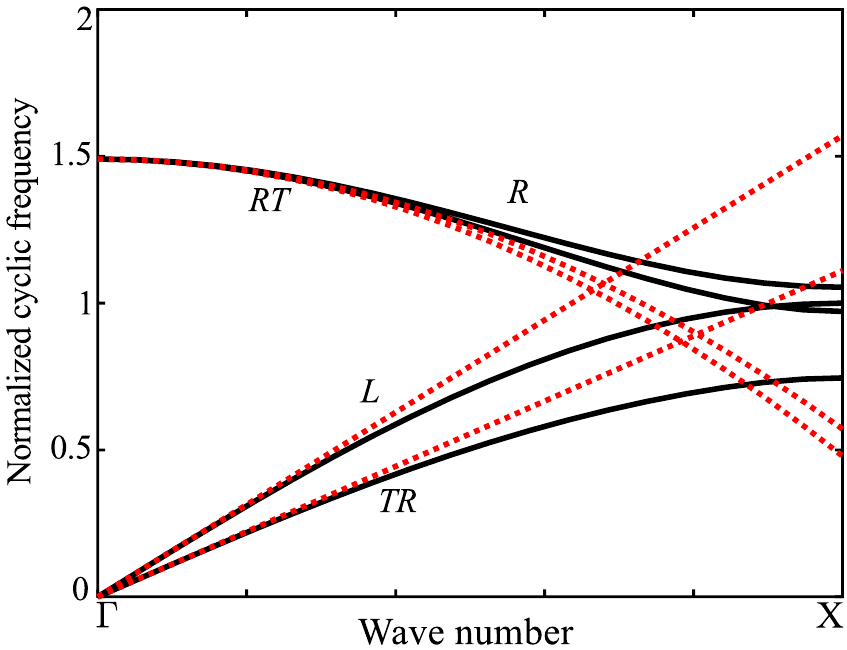}
\caption{(Color online) Dispersion curves in the $x$ direction of the frictional FCC structure from the discrete model (black curves) and their approximations (red dashed curves) with $\nu=0.3$, i.e. $\Delta_K\simeq 0.82$. The cyclic frequencies are normalized with the cutoff frequency of the longitudinal mode $\omega_L^c$.  }
\label{figdiscretedispersion}
\end{figure}

Considering only the stiffnesses $K_N$ and $K_S$ and the wave vector \textbf{k} in the $x$ direction, a pure longitudinal $L$ mode propagates, whose eigenvector is $\textbf{X}=(U_x,0,0,0,0,0)$ and has the dispersion relation
\begin{equation}
\omega_L^2=\frac{8}{m_b}(K_N+K_S) \sin^2\kappa,  
\label{eqdispnormal}
\end{equation}
where $\kappa=k_xa/(2\sqrt{2})$ is the normalized wavenumber, $\kappa=\pi/2$ is the minimal wavelength propagating. 
A pure rotational mode $R$ exists, whose eigenvector is $\textbf{X}=(0,0,0,W_x,0,0)$ and has the dispersion relation
\begin{equation}
\omega_R^2=\frac{20 K_S}{m_b}\biggl[1+\cos^2\kappa\biggl]. 	
\end{equation}
Two degenerate Transverse-Rotational (TR) and two degenerate Rotational-Transverse (RT) modes propagate, whose eigenvectors are $\textbf{X}=(0,U_y,0,0,0,W_z)$ and $\textbf{X}=(0,0,U_z,0,W_y,0)$ and have the dispersion relation
\begin{eqnarray}
\omega_{TR,RT}^2 & = & \frac{1}{m_b}\biggl\{ 2K_N\sin^2\kappa+9 K_S\cos^2\kappa +11K_S \nonumber \\
& & \pm \biggl[ 4K_N^2\sin^4\kappa+ \cos^2\kappa (399K_S^2-4K_NK_S) \nonumber\\
& & -  \sin^2 (2\kappa)(121K_S^2/4+21K_NK_S)+K_S^2+4K_NK_S\biggl]^{1/2}\biggl\},  
\end{eqnarray}
where the minus and plus signs correspond to the TR and RT modes, respectively. 
The small wavelength cut-off frequencies of all the modes for $\kappa=\pi/2$ are
\begin{eqnarray}
\omega_L^c & = & \sqrt{\frac{8(K_N+K_S)}{m_b}}=\sqrt{\frac{42(K_S+K_N)}{\pi \rho_b a^3}}, \label{eqcutoffL}\\
\omega_R^c & = & \sqrt{\frac{20K_S}{m_b}}=\sqrt{\frac{120K_S}{\pi\rho_ba^3}}, \\
\omega_{RT}^c & = & 2\sqrt{\frac{K_N+3K_S}{m_b}}=2\sqrt{\frac{6K_N+18K_S}{\pi\rho_ba^3}}, \\
\omega_{TR}^c & = & \sqrt{\frac{10 K_S}{m_b}}=\sqrt{\frac{60 K_S}{\pi\rho_b a^3}}.
\end{eqnarray}
The cut-off frequencies of the modes R and RT for $\kappa=0$ are
\begin{eqnarray}
\omega_{RT,R}(0) & = & \sqrt{\frac{40K_S}{m_b}}=\sqrt{\frac{240K_S}{\pi \rho_b a^3}}. 
\end{eqnarray}
It should be noticed that $\omega_{RT}(0)=\omega_{RT}(\kappa=\pi/2)$ if $K_S=K_N/7$, but $\omega_{RT}$ is not constant between these two points; the mode RT still propagates in this case. \\
For long wavelength (small wavenumber $\kappa=k_xa/(2\sqrt{2}) \ll \pi/2$), approximations of the dispersion relations can be found from a Taylor expansion
\begin{eqnarray}
\omega_L & \simeq & \sqrt{\frac{K_N+K_S}{m_b}}k_x a=\sqrt{\frac{6(K_N+K_S)}{\pi \rho_b a}}k_x, 
\label{eqdispdiscL} \\
\omega_{TR} & \simeq &  \sqrt{\frac{K_N+K_S}{2m_b}}k_x a = \sqrt{\frac{3(K_N+K_S)}{\pi \rho_b a}}k_x, 
\label{eqdispdiscTR} \\
\omega_{RT} & \simeq & \sqrt{\frac{40K_S}{m_b}}\biggl(1-\frac{11}{320}k_x^2 a^2\biggl) = \sqrt{\frac{240 K_S}{\pi \rho_b a^3}}\biggl(1-\frac{11}{320}k_x^2a^2\biggl), 
\label{eqdispdiscRT} \\
\omega_R & \simeq &  \sqrt{\frac{40K_S}{m_b}}\biggl(1-\frac{10}{320}k_x^2 a^2\biggl) = \sqrt{\frac{240K_S}{\pi \rho_b a^3}}\biggl(1-\frac{10}{320}k_x^2a^2\biggl). 
\label{eqdispdiscR}
\end{eqnarray} 
The scaled dispersion relations and their approximations are displayed in Fig. \ref{figdiscretedispersion} along the segment $\Gamma-$X of the first Brillouin zone. The coordinates of the points $\Gamma$ and X are $\Gamma=(0,0,0)$ and X$=(\pi\sqrt{2}/a,0,0)$ in the $(k_x,k_y,k_z)$-space. 
Modelling the contacts with a relation from the Hertz-Mindlin theory, the curves are determined only by the ratio of shear to longitudinal stiffness $\Delta_K$, which only depends on the Poisson's ratio of the material constituting the beads \cite{merkel2010PRE}. 


\subsection{Rolling and twisting resistant case}
\label{subsecdispdiscCOH}

\begin{figure}[ht!]
\centering
\includegraphics[width=14cm]{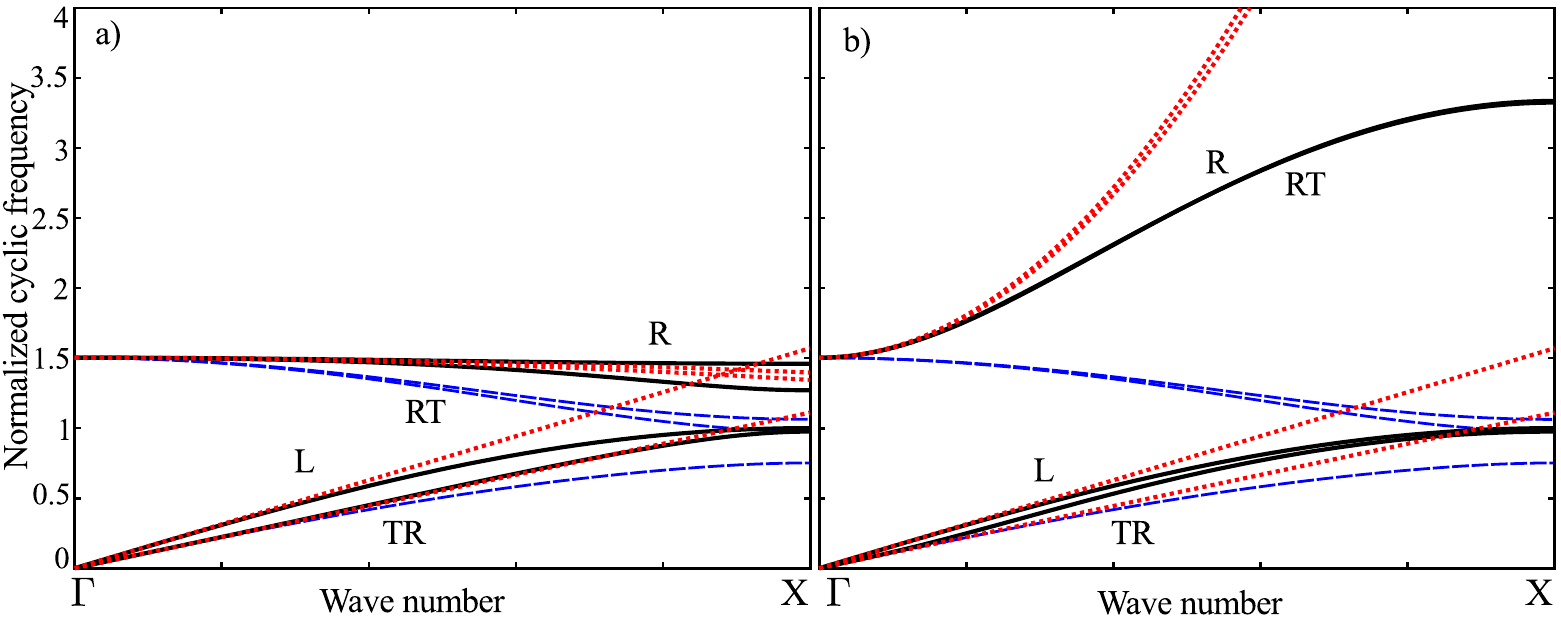}
\caption{(Color online) Dispersion curves in the $x$ direction of the rolling and twisting resistant FCC structure from the discrete model  (black continuous curves), their approximations (red dotted curves) and the dispersion curves of the frictional FCC structure from the discrete model (blue dashed curves) with $\Delta K=K_S/K_N \simeq 0.82$ and $a=2$~mm. (a) $G_B=K_Na^2/10$ and $G_T=K_Sa^2/10$. (b) $G_B=K_Na^2$ and $G_T=K_Sa^2$. The cyclic frequencies are normalized with the cut-off frequency of the longitudinal mode $\omega_L^c$. The rotational stiffnesses $G_B$ and $G_T$ are inspired by \cite{brendel2011}. 
}
\label{figdiscretedispersionCOH}
\end{figure}

Considering all the stiffnesses $K_N$, $K_S$, $G_T$, $G_B$ and the wave vector \textbf{k} in the $x$ direction, a longitudinal mode $L$ propagates, whose eigenvector is $\textbf{X}=(U_x,0,0,0,0,0)$ and has the dispersion relation
\begin{equation}
\omega_L^2=\frac{8}{m_b}(K_N+K_S) \sin^2\kappa,  
\end{equation}
which is identical to the dispersion relation in Eq. (\ref{eqdispnormal}) as expected. 
A pure rotational mode $R$ exists, whose eigenvector is $\textbf{X}=(0,0,0,W_x,0,0)$ and has the dispersion relation
\begin{equation}
\omega_R^2=\frac{20 K_S}{m_b}\biggl[1+\cos^2\kappa\biggl] +\frac{80(G_T+G_B)}{m_ba^2}\sin^2\kappa. 	
\end{equation}
Two degenerate transverse-rotational TR and two degenerate rotational-transverse RT modes propagate, whose eigenvectors are $\textbf{X}=(0,U_y,0,0,0,W_z)$ and $\textbf{X}=(0,0,U_z,0,W_y,0)$ and have the dispersion relation
\begin{eqnarray}
\omega_{TR,RT}^2 & = & \frac{1}{m_b}\biggl\{\biggl( \frac{60 G_B+20G_T}{a^2}+2K_N\biggl)\sin^2\kappa +9 K_S\cos^2\kappa +11K_S \nonumber\\
& \pm &\biggl[ 4\biggl(\frac{900G_B^2+100G_T^2}{a^4}+K_N^2-\frac{60K_NG_B}{a^2}+\frac{600G_TG_B}{a^4}-\frac{20K_NG_T}{a^2}\biggl)\sin^4\kappa\nonumber  \\
& + &K_S\biggl( \frac{630G_B+210G_T}{a^2}-21K_N\biggl)\sin^2(2\kappa)+4K_S \biggl( K_N-\frac{30G_B+10G_T}{a^2}\biggl)\sin^2\kappa \nonumber \\
& + & \frac{K_S^2}{4} \biggl( 121\cos^2( 2\kappa)+798\cos( 2\kappa)+681\biggl)\biggl]^{1/2}\biggl\},  
\label{eqdispcohRTTR}
\end{eqnarray}
where the minus and plus signs correspond to the TR and RT mode, respectively. 
The cutoff frequencies of all the modes for $\kappa=\pi/2$ can then be predicted as
\begin{eqnarray}
\omega_L^c & = & \sqrt{8\frac{K_N+K_S}{m_b}}=\sqrt{42\frac{K_S+K_N}{\pi \rho_b a^3}}, \\
\omega_R^c & = & \sqrt{20\frac{K_S+4(G_T+G_B)/a^2}{m_b}}=\sqrt{120\frac{K_S+4(G_T+G_B)/a^2}{\pi\rho_ba^3}}, \label{eqcutoffR}\\
\omega_{TR,RT}^c & = & \sqrt{10\frac{K_Sa^2+12G_B+4G_T}{a^2m_b}}=\sqrt{60\frac{K_S+(12G_B+4G_T)/a^2}{\pi \rho_b a^3}}, \label{eqcutoffRTTR1}\\
\omega_{TR,RT}^c & = &2\sqrt{\frac{K_N+3K_S}{m_b}}=2\sqrt{6\frac{K_N+3K_S}{\pi\rho_ba^3}}. \label{eqcutoffRTTR2}
\end{eqnarray}
If $\kappa=\pi/2$, Eq. (\ref{eqdispcohRTTR}) reduces to Eqs. (\ref{eqcutoffRTTR1}) or (\ref{eqcutoffRTTR2}) without distinction whether the sign considered is plus or minus. To discriminate between the two cutoff frequencies, the repulsion property between the mode TR and RT may be used. If Eq. (\ref{eqcutoffRTTR1})$>$Eq. (\ref{eqcutoffRTTR2}), then Eq. (\ref{eqcutoffRTTR1}) is the cutoff frequency of the mode RT and vice versa. 
The cutoff frequencies of the modes R and RT for $\kappa=0$ can then be predicted as
\begin{equation}
\omega_{RT,R}(0) = \sqrt{\frac{40K_S}{m_b}}=\sqrt{\frac{240K_S}{\pi \rho_b a^3}}.
\label{eqcutoffkzerocoh} 
\end{equation}
For long wavelength (small wavenumber $\kappa=k_xa/(2\sqrt{2}) \ll \pi/2$), the approximations of the dispersion relations can be found from a Taylor expansion
\begin{eqnarray}
\omega_L & \simeq & \sqrt{\frac{K_N+K_S}{m_b}}k_x a=\sqrt{\frac{6(K_N+K_S)}{\pi \rho_b a}}k_x, 
\label{eqdispdiscLCOH} \\
\omega_{TR} & \simeq &  \sqrt{\frac{K_N+K_S}{2m_b}}k_x a = \sqrt{\frac{3(K_N+K_S)}{\pi \rho_b a}}k_x, 
\label{eqdispdiscTRCOH} \\
\omega_{RT} & \simeq  &\sqrt{\frac{40K_S}{m_b}}\biggl(1+\frac{3G_B+G_T}{16K_S}k_x^2-\frac{11a^2}{320}k_x^2 \biggl) \nonumber \\
 & = & \sqrt{\frac{240 K_S}{\pi \rho_b a^3}}\biggl(1+\frac{3G_B+G_T}{16K_S}k_x^2-\frac{11a^2}{320}k_x^2\biggl), 
\label{eqdispdiscRTCOH} \\
\omega_R & \simeq  & \sqrt{\frac{40K_S}{m_b}}\biggl(1+\frac{G_B+G_T}{8K_S}k_x^2-\frac{10a^2}{320}k_x^2\biggl) \nonumber \\
 &=  &\sqrt{\frac{240K_S}{\pi \rho_b a^3}}\biggl(1+\frac{G_B+G_T}{8K_S}k_x^2-\frac{10a^2}{320}k_x^2\biggl). 
\label{eqdispdiscRCOH}
\end{eqnarray} 
In the rolling and twisting resistant case, the dispersion curves depend on the 4 stiffness $K_N,\:K_S,\:G_T\:,G_B$ and the diameter of the beads $a$. In Fig. \ref{figdiscretedispersionCOH}, the dispersion relations and their approximations are displayed in two different cases. In the first one, the cutoff frequencies of the modes R and RT at $\kappa=\pi/2$ are smaller than the ones at $\kappa=0$. In the second case, the situation is opposite. 

In conclusion of this section, the long wavelength approximation of the discrete model is derived in the frictional (sliding resistant) case in Eqs. (\ref{eqdispdiscL})-(\ref{eqdispdiscR}) and in the rolling and twisting resistant case in Eqs. (\ref{eqdispdiscLCOH})-(\ref{eqdispdiscRCOH}). These equations are considered as references in the comparison with the continuum models that are derived in the following sections. 


\section{Numerical wave propagation}
\label{secsimu}

In this section, in order to validate the theoretical predictions of the discrete model,
the wave propagation is numerically simulated \cite{mouraille2006} with a packing arranged 
in a FCC structure of equal sized beads. The crystal is formed by square cubic layers of $4\times4$ particles 
in the $y-z$ plane, which are stacked in the closest position to each other in the $x$ direction in an ABAB... 
sequence, where the B layers are shifted in the $y-z$ plane to be able to fill the deepest holes in the A layers. 
The crystal is composed of 200 layers, giving a total of 3200 beads, while periodic boundary conditions are 
considered in all directions. 
The beads have a diameter $a=2$~mm and are composed of glass with a mass density $\rho_b=2000$~kg.m$^{-3}$, 
a Young modulus $E=69$~GPa, and a Poisson's ratio $\nu=0.3$. 
The normal static force applied on each contact in the initial configuration is $F_N^0=2.10^{-3}$~N, 
which, according to the Hertz law \cite{merkel2010PRE,duffy1956,Bjohnson1985,thornton1991,gilles2003} where 
\begin{equation}
K_{N}=(3aF_N^0/8)^{\frac{1}{3}}E^{\frac{2}{3}}(1-\nu^2)^{-\frac{2}{3}}, 
\end{equation}
and
\begin{equation}
\delta_0=2\biggl[\frac{3}{4}\frac{(1-\nu^2)F_N^0}{E\sqrt{a/2}}\biggl]^{2/3},
\end{equation}
gives a normal stifness $K_N=2.051.10^{5}$~N.m$^{-1}$ and an initial inter-particle relative displacement 
$\delta_0=1.463.10^{-8}$~m. 
The wave is excited on a single layer (here, the second layer) by imposing, depending on the mode to excite, a velocity, a rotation angle and a rotation/angular velocity.
To excite the L mode (planar compressive P-wave) of propagation in $x$-direction, 
all the particles in the second layer get a (small) velocity $v_x^0$. In contrast, 
to excite the TR and RT modes (shear-, or S-waves) of propagation in $x$-direction, 
all the particles in the second layer get a (small) velocity $v_y^0$ or $v_z^0$. 
Finally, to excite the R mode, all the particles of the second layer get a rotation/angular 
velocity $w_x^0$ and a rotation angle $q^0_x$, both parallel to the $x$ direction of propagation. 
The dispersion relations are found by computing a two-dimensional Fast Fourier Transform (FFT) 
in time and space using the velocities and angles per layer, sampled in space at every layer
and in time at every $\Delta t = 10^{-6}$~s with 2048~points.

\begin{figure}[ht!]
\centering
\includegraphics[width=15cm]{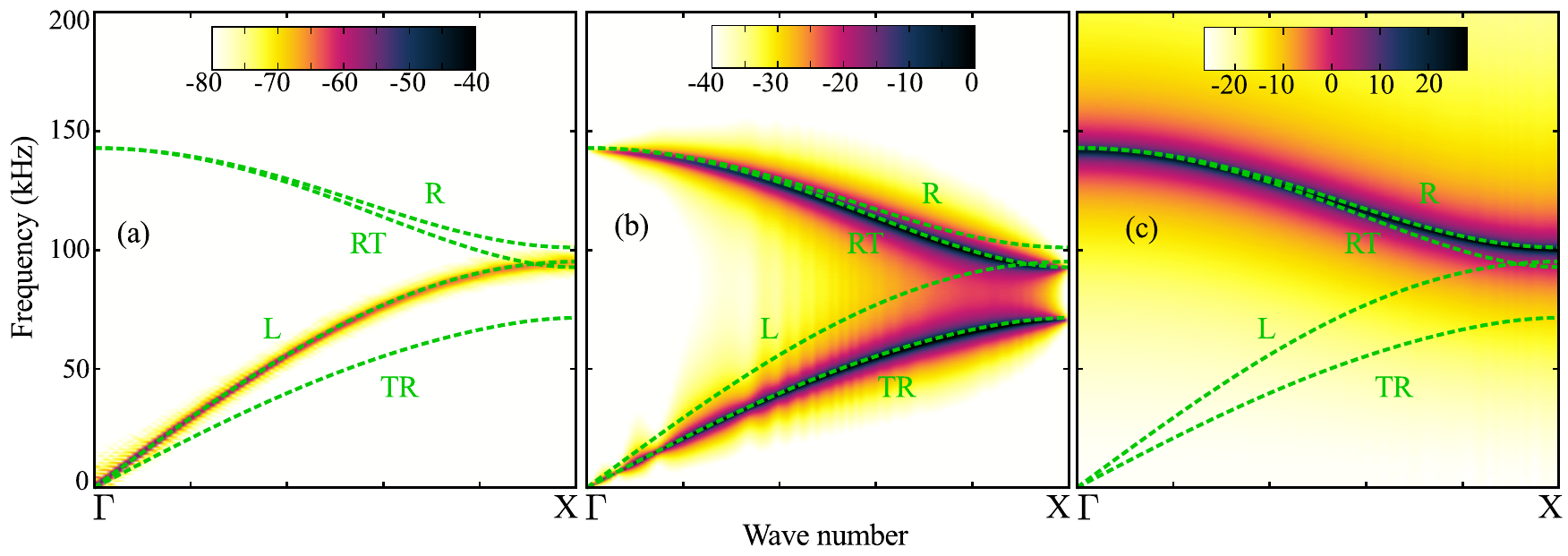}
\caption{(Color online) Dispersion relations from numerical wave propagation in the FCC structure with normal stiffness
$K_N=2.051.10^5$~N.m$^{-1}$ and tangential stiffness $K_S=\Delta_K K_N=1.689.10^5$~N.m$^{-1}$,
i.e. $\Delta_K \approx 0.82$, where the inset shows the magnitude of the two-dimensional FFT in a logscale with arbitrary reference. 
The dispersion relations from the discrete mode (dahsed green curves) correspond to the solid lines in 
Fig.\ \ref{figdiscretedispersion}, with activated (a) L mode, (b) TR and RT modes, and (c) R mode. Note
that in contrast to Fig. \ref{figdiscretedispersion}, the vertical axis is not scaled. 
}
\label{figSimNonCoh}
\end{figure}
\begin{figure}[ht!]
\centering
\includegraphics[width=8cm]{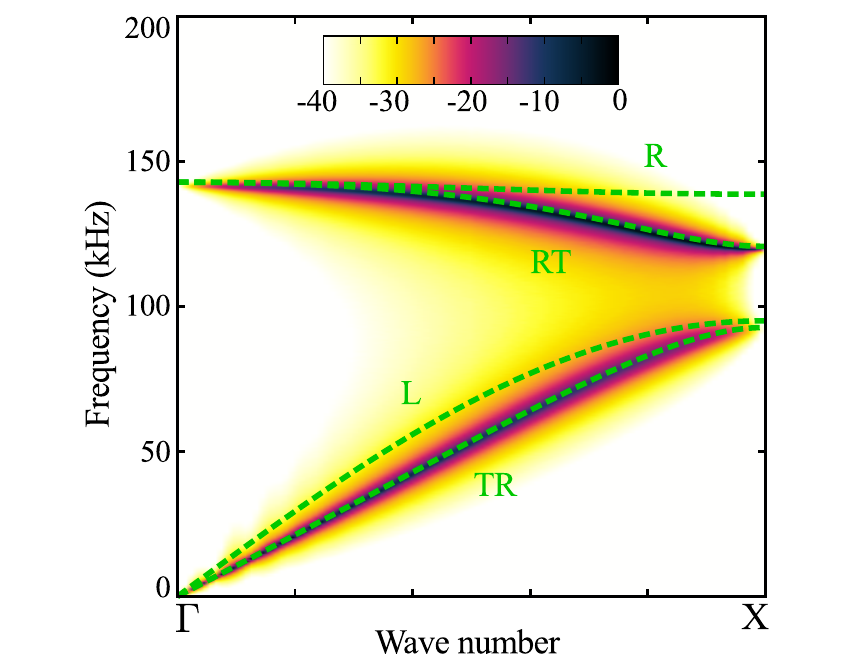}
\caption{(Color online) Dispersion relations from numerical wave propagation in the rolling and twisting resistant FCC structure 
with $K_N=2.051.10^5$~N.m$^{-1}$, $K_S=\Delta_K K_N=1.689.10^5$~N.m$^{-1}$, 
$G_B=K_N a^2/10=8.20.10^{-2}$~N.m/rad and $G_T=K_S a^2/10=6.76.10^{-2}$~N.m/rad,
with the magnitude of the two-dimensional FFT in logscale with arbitrary reference. Dispersion relations (green curves) from the discrete model correspond to the solid lines in Fig. \ref{figdiscretedispersionCOH}. 
Only the TR and RT modes are shown. The dispersion relation of the L mode is identical to the one of the frictional case. The R mode is very slowly propagating (almost flat dispersion curve in this case) and requires a much longer time of simulation. 
}
\label{figSimCoh}
\end{figure}

The dispersion relations for the ''frictional'' case with tangential stiffness\footnote{The tangential elasticity $K_S$ is active and the coefficient of friction is set to a very large 
value in order to avoid slip at the contacts and such as to ensure the propagation of an elastic wave. } 
are plotted in Fig. \ref{figSimNonCoh}. Depending on the different types of excitation, the dispersion relations 
reflect the respective modes and the numerical simulations correspond exactly to those of all the modes predicted 
by the discrete model. 

The dispersion relations of the TR and RT modes in the rolling and twisting resistant case are plotted 
in Fig. \ref{figSimCoh} and also show and confirm the good agreement between the numerical simulation results and the 
predictions of the discrete model. 
It should be noticed that the complete dispersion curves are found from one single numerical simulation,
since the agitation is a discontinuous change in velocity and thus activates all frequencies, which
leads to a very clean signal for the perfect granular crystal structure used here. 
The introduction of, even weak, polydispersity in the diameters of the beads affects the wave propagation and, 
as a result, only the small wave number part of the dispersion relation (long wavelengths) can be retrieved 
\cite{mouraille2008}. 

In summary, these simulations confirm/validate the discrete model, on which we continue 
by simplifying and trying to identify a simpler model for large wavelengths.

\section{From the discrete description to the continuum description}
\label{seccontinuum}

In this section, the macroscopic stress and couple stress tensors are derived through a micro-macro transition of the applied force and torque between two beads of the discrete model in Eqs. (\ref{eqforce}) and (\ref{eqmoment}). It follows the derivations in \cite{suiker2001-1}. In Sec. \ref{subsecgradient}, the discrete displacements and rotations are transformed into continuum fields. The order of the Taylor expansion is chosen to derived a second-order gradient micropolar model. In Sec. \ref{subsecforce}, the displacement gradient is inserted into the force equation in Eq. (\ref{eqforce}). In Sec. \ref{subsecmoment}, the particle rotation is inserted into the torque equation in Eq. (\ref{eqmoment}). In Sec. \ref{subseccauchystress}, the Cauchy stress tensor is derived. In Sec. \ref{subseccouplestress}, the couple stress tensor is derived. The Cauchy stress and the couple stress tensors form the constitutive equation of the second-order gradient micropolar model. 
Finally in Sec. \ref{subsecConsitutive}, the constitutive equations of the classical, second-order gradient and Cosserat model of elasticity are retrieved from the second-order gradient micropolar model by setting some of the tensors of this latter model to zero. 

\subsection{Expansion of the displacement and angular rotation}
\label{subsecgradient}

The discrete displacements and rotations are transformed into continuum fields using a Taylor expansion (the choice of the order of the expansion will be explained below)
\begin{eqnarray}
u_i^{\beta} &=& u_i^{\alpha}+D_j^{\beta\alpha}\frac{\partial u_i^{\alpha}}{\partial x_j}+\frac{1}{2}D_j^{\beta\alpha}D_k^{\beta\alpha}\frac{\partial^2 u_i^{\alpha}}{\partial x_j\partial x_k}+\frac{1}{6}D_j^{\beta\alpha}D_k^{\beta\alpha}D_l^{\beta\alpha}\frac{\partial^3 u_i^{\alpha}}{\partial x_j\partial x_k\partial x_l}+\cdots\nonumber\\
w_i^{\beta} &=& w_i^{\alpha}+D_j^{\beta\alpha}\frac{\partial w_i^{\alpha}}{\partial x_j}+\frac{1}{2}D_j^{\beta\alpha}D_k^{\beta\alpha}\frac{\partial^2 w_i^{\alpha}}{\partial x_j\partial x_k}+\cdots
\label{eqgradient}
\end{eqnarray}
The displacement gradient $\partial u^{\alpha}_i/\partial x_j=u^{\alpha}_{\langle i,j\rangle}+u^{\alpha}_{[i,j]}$ can be decomposed into its symmetric part (which is the classical strain)
\begin{eqnarray}
u^{\alpha}_{\langle i,j\rangle} & = & \epsilon^{\alpha}_{ij}=\epsilon^{\alpha}_{ji}=\frac{1}{2}\biggl(\frac{\partial u^{\alpha}_i}{\partial x_j}+\frac{\partial u^{\alpha}_j}{\partial x_i}\biggl),
\end{eqnarray}
and its antisymmetric part
\begin{eqnarray}
u^{\alpha}_{[i,j]} & = & -u^{\alpha}_{[j,i]}=\frac{1}{2}\biggl(\frac{\partial u^{\alpha}_i}{\partial x_j}-\frac{\partial u^{\alpha}_j}{\partial x_i}\biggl). 
\label{eqidu}
\end{eqnarray}
The particle rotation 
\begin{equation}
w_i^{\alpha}=\Pi_i^{\alpha}+\Gamma_i^{\alpha}
\label{eqidw}
\end{equation}
 is composed of two components, the particle spin $\Pi_i^{\alpha}$ and the vorticity $\Gamma_i^{\alpha}$. The particle spin is independent of the displacement field, whereas the vorticity is completely generated by the antisymmetric part of the displacement gradient according to 
\begin{equation}
\Gamma^{\alpha}_i=-\frac{1}{2}\varepsilon_{ijk}u^{\alpha}_{[j,k]}
\label{eqsolidrot}
\end{equation} 
and reciprocally
\begin{equation}
u^{\alpha}_{[i,j]}=-\varepsilon_{ijk} \Gamma^{\alpha}_k. 
\label{eqsolidrot2}
\end{equation}
The vorticity is here the local rotation of the assembly of particles, i.e., the rigid body rotation or the background rotation. 
In order to have the same order $c$ of derivatives in the model, the order of derivative of $\textbf{u}$ should be one order higher than the order of derivative of $\textbf{w}$ as $\{c,c-1\}$ as detailed in \cite{suiker2001-1}. In order to get the second-order gradient micropolar constitutive equations, the order of derivatives is chosen to be $\{3,2\}$. 

\subsection{Relative displacement and applied force}
\label{subsecforce}

Putting Eq. (\ref{eqgradient}) in the relative displacement $\delta^{\beta\alpha}_i$ in Eq. (\ref{eqforce}), yields
\begin{eqnarray}
\delta_i^{\beta\alpha} & = & D_j^{\beta\alpha}\frac{\partial u^{\alpha}_i}{\partial x_j}+\frac{1}{2}D_j^{\beta\alpha}D_k^{\beta\alpha}\frac{\partial^2 u_i^{\alpha}}{\partial x_j\partial x_k}+\frac{1}{6}D_j^{\beta\alpha}D_k^{\beta\alpha}D_l^{\beta\alpha}\frac{\partial^3 u_i^{\alpha}}{\partial x_j\partial x_k\partial x_l}\nonumber\\
& + & \frac{1}{2} \varepsilon_{ijk}D_j^{\beta\alpha} (2w^{\alpha}_k+D_l^{\beta\alpha}\frac{\partial w^{\alpha}_k}{\partial x_l}+\frac{1}{2}D_l^{\beta\alpha}D_p^{\beta\alpha}\frac{\partial^2 w^{\alpha}_k}{\partial x_l\partial x_p}). 
\label{eqdiscont}
\end{eqnarray}
Inserting the identities (\ref{eqidu}) and (\ref{eqidw}), Eq. (\ref{eqdiscont}) becomes 
\begin{eqnarray}
\delta_i^{\beta\alpha} & = &  D_j^{\beta\alpha} (\epsilon^{\alpha}_{ij}+u^{\alpha}_{[i,j]}) + \frac{1}{2}D_j^{\beta\alpha}D_k^{\beta\alpha}\biggl(\frac{\partial \epsilon^{\alpha}_{ij}}{\partial x_k}+\frac{\partial u^{\alpha}_{[i,j]}}{\partial x_k}\biggl)+\frac{1}{6}D_j^{\beta\alpha}D_k^{\beta\alpha}D_l^{\beta\alpha}\biggl(\frac{\partial^2 \epsilon^{\alpha}_{ij}}{\partial x_k\partial x_l}+\frac{\partial^2 u^{\alpha}_{[i,j]}}{\partial x_k \partial x_l}\biggl) \nonumber \\
& + & \frac{1}{2} \varepsilon_{ijk}D_j^{\beta\alpha} \biggl[ 2(\Pi^{\alpha}_k+\Gamma^{\alpha}_k) +D_l^{\beta\alpha} \biggl( \frac{\partial \Pi^{\alpha}_k}{\partial x_l}+ \frac{\partial\Gamma^{\alpha}_k}{\partial x_l}\biggl)+ \frac{1}{2}D_l^{\beta\alpha}D_p^{\beta\alpha} \biggl( \frac{\partial^2 \Pi^{\alpha}_k}{\partial x_l\partial x_p}+ \frac{\partial^2\Gamma^{\alpha}_k}{\partial x_l\partial x_p}\biggl)\biggl]. \nonumber\\
\end{eqnarray}
With the help of Eqs. (\ref{eqsolidrot}) and (\ref{eqsolidrot2}), it becomes
\begin{eqnarray}
\delta_i^{\beta\alpha} & = &  D_j^{\beta\alpha} \epsilon^{\alpha}_{ij} + \frac{1}{2}D_j^{\beta\alpha}D_k^{\beta\alpha}\frac{\partial \epsilon^{\alpha}_{ij}}{\partial x_k}+\frac{1}{6}D_j^{\beta\alpha}D_k^{\beta\alpha}D_l^{\beta\alpha}\frac{\partial^2 \epsilon^{\alpha}_{ij}}{\partial x_k\partial x_l}-\frac{1}{12} D_j^{\beta\alpha}D_k^{\beta\alpha}D_l^{\beta\alpha}\frac{\partial^2 u^{\alpha}_{[i,j]}}{\partial x_k\partial x_l}\nonumber \\
& + & \frac{1}{2} \varepsilon_{ijk}D_j^{\beta\alpha} \biggl[ 2\Pi^{\alpha}_k+D_l^{\beta\alpha} \frac{\partial \Pi^{\alpha}_k}{\partial x_l}+ \frac{1}{2}D_l^{\beta\alpha}D_p^{\beta\alpha}  \frac{\partial^2 \Pi^{\alpha}_k}{\partial x_l\partial x_p}\biggl]. 
\label{eqdisfin}
\end{eqnarray}
From Eqs. (\ref{eqdisfin}) and (\ref{eqforce}), the force applied by a bead $\beta$ on a bead $\alpha$ is written as
\begin{eqnarray}
F^{\beta\alpha}_i & = & K^{\beta\alpha}_{ij}\delta_j^{\beta\alpha} =  K^{\beta\alpha}_{ij}D_k^{\beta\alpha} \epsilon^{\alpha}_{jk} + \frac{1}{2}K^{\beta\alpha}_{ij}D_k^{\beta\alpha}D_l^{\beta\alpha}\frac{\partial \epsilon^{\alpha}_{jk}}{\partial x_l}+ \frac{1}{6}K^{\beta\alpha}_{ij}D_k^{\beta\alpha}D_l^{\beta\alpha}D_m^{\beta\alpha}\frac{\partial^2 \epsilon^{\alpha}_{kj}}{\partial x_l\partial x_m} \nonumber \\
& - &\frac{1}{12} K^{\beta\alpha}_{ij}D_k^{\beta\alpha}D_l^{\beta\alpha}D_m^{\beta\alpha}\frac{\partial^2 u^{\alpha}_{[j,k]}}{\partial x_l\partial x_m} +  K^{\beta\alpha}_{ij}\varepsilon_{jkl}D_k^{\beta\alpha} \Pi^{\alpha}_k +\frac{1}{2} K^{\beta\alpha}_{ij}\varepsilon_{jkl}D_k^{\beta\alpha}D_m^{\beta\alpha} \frac{\partial \Pi^{\alpha}_l}{\partial x_m} \nonumber \\
& + & \frac{1}{4} K^{\beta\alpha}_{ij}\varepsilon_{jkl}D_k^{\beta\alpha}D_m^{\beta\alpha}D_p^{\beta\alpha}  \frac{\partial^2 \Pi^{\alpha}_l}{\partial x_m\partial x_p}. 
\label{eqforcecont}
\end{eqnarray}

\subsection{Relative angular rotation and applied torque}
\label{subsecmoment}

Inserting Eq. (\ref{eqgradient}) in the relative angular rotation $\theta^{\beta\alpha}_j$ in Eq. (\ref{eqmoment}), the relative angular rotation becomes
\begin{equation}
\theta^{\beta\alpha}_i=D^{\beta\alpha}_k\frac{\partial w^{\alpha}_j}{\partial x_k} + \frac{1}{2}D^{\beta\alpha}_kD^{\beta\alpha}_l\frac{\partial^2 w^{\alpha}_j}{\partial x_k\partial x_l}. 
\label{eqrotcont}
\end{equation}
Inserting Eqs. (\ref{eqidw}) and (\ref{eqsolidrot}) in Eq. (\ref{eqrotcont}), the applied torque in Eq. (\ref{eqmoment}) becomes
\begin{eqnarray}
M^{\beta\alpha}_i = G_{ij} \biggl( D^{\beta\alpha}_k\frac{\partial \Pi^{\alpha}_j}{\partial x_k} + \frac{1}{2}D^{\beta\alpha}_kD^{\beta\alpha}_l\frac{\partial^2 \Pi^{\alpha}_j}{\partial x_k\partial x_l} +  \frac{1}{2}\epsilon_{jnm}D^{\beta\alpha}_k\frac{\partial u^{\alpha}_{[m,n]}}{\partial x_k} + \frac{1}{4}\epsilon_{jnm}D^{\beta\alpha}_kD^{\beta\alpha}_l\frac{\partial^2 u^{\alpha}_{[m,n]}}{\partial x_k\partial x_l}\biggl). \nonumber\\
\label{eqmomentcont}
\end{eqnarray}

\subsection{Cauchy stress tensor}
\label{subseccauchystress}

The Cauchy stress tensor for a particle $\alpha$ in contact with $N$ other particles $\beta$ can be written as \cite{suiker2001-1}
\begin{eqnarray}
\sigma_{ij}  & = & \frac{1}{V_{cell}} \sum_{\beta=1}^{N} F_j^{\beta\alpha}D_i^{\beta\alpha},  
\label{eqstresscauchy} 
\end{eqnarray}
where $V_{cell}$ is the volume of the unit cell which is composed of two beads, see Sec. \ref{secCosserat}. 
Combining Eqs. (\ref{eqforcecont}) and (\ref{eqstresscauchy}), the stress tensor can be expressed as
\begin{eqnarray}
\sigma_{hi} & = & \frac{1}{V_{cell}} \sum_{\beta=1}^{N}\biggl[ K^{\beta\alpha}_{ij}D_k^{\beta\alpha} D_h^{\beta\alpha}\epsilon^{\alpha}_{jk} + \frac{1}{2}K^{\beta\alpha}_{ij}D_k^{\beta\alpha}D_l^{\beta\alpha}D_h^{\beta\alpha}\frac{\partial \epsilon^{\alpha}_{jk}}{\partial x_l}+ \frac{1}{6}K^{\beta\alpha}_{ij}D_k^{\beta\alpha}D_l^{\beta\alpha}D_m^{\beta\alpha}D_h^{\beta\alpha}\frac{\partial^2 \epsilon^{\alpha}_{kj}}{\partial x_l\partial x_m} \nonumber \\
& - &\frac{1}{12} K^{\beta\alpha}_{ij}D_k^{\beta\alpha}D_l^{\beta\alpha}D_m^{\beta\alpha}D_h^{\beta\alpha}\frac{\partial^2 u^{\alpha}_{[j,k]}}{\partial x_l\partial x_m} +  K^{\beta\alpha}_{ij}\varepsilon_{jkl}D_k^{\beta\alpha} D_h^{\beta\alpha}\Pi^{\alpha}_k +\frac{1}{2} K^{\beta\alpha}_{ij}\varepsilon_{jkl}D_k^{\beta\alpha}D_m^{\beta\alpha} D_h^{\beta\alpha}\frac{\partial \Pi^{\alpha}_l}{\partial x_m} \nonumber \\
& + & \frac{1}{4} K^{\beta\alpha}_{ij}\varepsilon_{jkl}D_k^{\beta\alpha}D_m^{\beta\alpha}D_p^{\beta\alpha}D_h^{\beta\alpha} \frac{\partial^2 \Pi^{\alpha}_l}{\partial x_m\partial x_p} \biggl], \nonumber \\
\sigma_{ij} & = & C_{ijkl}\epsilon_{kl}+D_{ijklm}\frac{\partial \epsilon_{kl}}{\partial x_m}+E_{ijklmn}\frac{\partial^2 \epsilon_{kl}}{\partial x_m\partial x_n}+F_{ijklmn}\frac{\partial^2 u_{[k,l]}}{\partial x_m\partial x_n}\nonumber\\
& + & M_{ijm}\Pi_m+N_{ijmp}\frac{\partial \Pi_m}{\partial x_p}+P_{ijmpq}\frac{\partial^2 \Pi_m}{\partial x_p\partial x_q}. 
\end{eqnarray}
The constitutive tensors are therefore defined as
\begin{eqnarray}
C_{ijkl} & = & \frac{1}{V_{cell}} \sum_{\beta=1}^{N} K^{\beta\alpha}_{jk}D_l^{\beta\alpha} D_i^{\beta\alpha}, 
\label{eqC} \\
D_{ijklm} & = & \frac{1}{2V_{cell}} \sum_{\beta=1}^{N} K^{\beta\alpha}_{jk}D_l^{\beta\alpha}D_m^{\beta\alpha}D_i^{\beta\alpha},\\
E_{ijklmn} & = & \frac{1}{6V_{cell}} \sum_{\beta=1}^{N} K^{\beta\alpha}_{jk}D_l^{\beta\alpha}D_m^{\beta\alpha}D_n^{\beta\alpha}D_i^{\beta\alpha}, \\
F_{ijklmn} & = & -\frac{1}{12V_{cell}} \sum_{\beta=1}^{N} K^{\beta\alpha}_{jk}D_l^{\beta\alpha}D_m^{\beta\alpha}D_n^{\beta\alpha}D_i^{\beta\alpha}, \\
M_{ijm} & = & \frac{1}{V_{cell}} \sum_{\beta=1}^{N}K^{\beta\alpha}_{jk}\varepsilon_{klm}D_l^{\beta\alpha} D_i^{\beta\alpha}, 
\label{eqM} \\
N_{ijmp} & = & \frac{1}{2V_{cell}} \sum_{\beta=1}^{N} K^{\beta\alpha}_{jk}\varepsilon_{klm}D_l^{\beta\alpha}D_p^{\beta\alpha} D_i^{\beta\alpha},\\
P_{ijmpq} & = & \frac{1}{4V_{cell}} \sum_{\beta=1}^{N} K^{\beta\alpha}_{jk}\varepsilon_{klm}D_l^{\beta\alpha}D_p^{\beta\alpha}D_q^{\beta\alpha}D_i^{\beta\alpha}. 
\label{eqP}
\end{eqnarray}
For a statistically homogeneous material, which is assumed to be the case for granular materials, the constitutive tensors have to be centrally symmetrical \cite{suiker2001-1,suiker2005,chang1995IJSS,dellisola2009}; the tensor should have an even order. The tensors $D_{ijklm}$ and $N_{ijkl}$ do not behave centrally symmetrical, thus we set $D_{ijklm}=N_{ijkl}=0$, so that the stress tensor reduces to
\begin{eqnarray}
\sigma_{ij}=C_{ijkl}\epsilon_{kl}+E_{ijklmn}\frac{\partial^2 \epsilon_{kl}}{\partial x_m\partial x_n}+F_{ijklmn}\frac{\partial^2 u_{[k,l]}}{\partial x_m\partial x_n}+M_{ijm}\Pi_m+P_{ijmpq}\frac{\partial^2 \Pi_m}{\partial x_p\partial x_q}. 
\label{eqSGstress}
\end{eqnarray}
The different elements of this stress tensor can be related to the different interactions between the grains. The tensors $C_{ijkl}$ and $E_{ijklmn}$ are related to the classical translational normal and shear interactions with the first order and second order gradient of displacement, respectively. The tensor $F_{ijklmn}$ is related to the gradient of vorticity that participate to the rotational contribution to the translational surface shear interaction through the sliding resistance, i.e., the rotation activate the shear stiffness $K_S$. The tensors $M_{ijm}$ and $P_{ijmpq}$ are related to the particle rotation interaction with the first order and second order gradient rotation, respectively, and participate to the rotational contribution to the translational surface shear interaction through the sliding resistance. 

\subsection{Couple stress tensor}
\label{subseccouplestress}

Following the reasoning of the Cauchy stress tensor, the couple stress tensor can be written as \cite{suiker2001-1}
\begin{equation}
\mu_{ij}  =  \frac{1}{V_{cell}} \sum_{\beta=1}^{N} M_j^{\beta\alpha}D_i^{\beta\alpha}. 
\label{eqcouplecauchy}
\end{equation}
Combining Eqs. (\ref{eqmomentcont}) and (\ref{eqcouplecauchy}), the couple stress tensor is expressed as 
\begin{eqnarray}
\mu_{ij} & = & \frac{1}{V_{cell}} \sum_{\beta=1}^{N} \biggl[G_{jk} D^{\beta\alpha}_lD_i^{\beta\alpha}\frac{\partial \Pi^{\alpha}_k}{\partial x_l} + \frac{1}{2}G_{jk}D^{\beta\alpha}_lD^{\beta\alpha}_pD_i^{\beta\alpha}\frac{\partial^2 \Pi^{\alpha}_k}{\partial x_l\partial x_p}  \nonumber \\
& + &\frac{1}{2}G_{jk}\epsilon_{knm}D^{\beta\alpha}_lD_i^{\beta\alpha}\frac{\partial u^{\alpha}_{[m,n]}}{\partial x_l} 
 +  \frac{1}{4}G_{jk}\epsilon_{knm}D^{\beta\alpha}_lD^{\beta\alpha}_pD_i^{\beta\alpha}\frac{\partial^2 u^{\alpha}_{[m,n]}}{\partial x_l\partial x_p}  \biggl], \\
 & = & Q_{ijkl} \frac{\partial \Pi_k}{\partial x_l}+R_{ijklp}\frac{\partial^2 \Pi_k}{\partial x_l \partial x_p} + S_{ijnml} \frac{\partial u_{[m,n]}}{\partial x_l}+ T_{ijnmlp} \frac{\partial u_{[m,n]}}{\partial x_l \partial x_p}. 
\end{eqnarray}
The constitutive tensors are therefore defined with
\begin{eqnarray}
Q_{ijkl} & = & \frac{1}{V_{cell}} \sum_{\beta=1}^{N} G_{jk} D^{\beta\alpha}_lD_i^{\beta\alpha},  \label{eqQtensorFCC}\\
R_{ijklp} & = &\frac{1}{2V_{cell}} \sum_{\beta=1}^{N} G_{jk}D^{\beta\alpha}_lD^{\beta\alpha}_pD_i^{\beta\alpha}, \\
S_{ijnml} & = & \frac{1}{2V_{cell}} \sum_{\beta=1}^{N} G_{jk}\epsilon_{knm}D^{\beta\alpha}_lD_i^{\beta\alpha}, \label{eqStensorFCC} \\ 
T_{ijnmlp} & = & \frac{1}{4V_{cell}} \sum_{\beta=1}^{N} G_{jk}\epsilon_{knm}D^{\beta\alpha}_lD^{\beta\alpha}_pD_i^{\beta\alpha}.  
\end{eqnarray}
Similarly to the case of the tensors $D_{ijklm}$ and $N_{ijkl}$, the tensors $R_{ijklp}$ and $T_{ijnmlp}$ do not behave centrally symmetrical, thus  $R_{ijklp}=T_{ijnmlp}=0$. The couple stress tensor then becomes
\begin{equation}
\mu_{ij} = Q_{ijkl} \frac{\partial \Pi_k}{\partial x_l} + S_{ijnml} \frac{\partial u_{[m,n]}}{\partial x_l}. 
\label{eqSGcouplestress}
\end{equation}
The tensor $Q_{ijkl}$ is related to the particle rotation interaction that activates both twisting and rolling resistances. The tensor $S_{ijnml}$ is related to the gradient of vorticity and also activates both twisting and rolling resistances. 

\subsection{Relation between the classical, second order gradient and Cosserat models}
\label{subsecConsitutive}

The stress tensor in Eq. (\ref{eqSGstress}) and the couple stress tensor in Eq. (\ref{eqSGcouplestress}) can be compared to those of others elastic models. 
The tensor $C_{ijkl}$ refers to the classical elastic tensor. In order to take into account all the symmetries of the tensor of deformation, the tensor $C_{ijkl}$ can be determined with \cite{ashcroft1976}
\begin{equation}
C_{ijkl}=\frac{1}{4V_{cell}} \sum_{\beta=1}^{N} K^{\beta\alpha}_{jk}D_l^{\beta\alpha} D_i^{\beta\alpha}+ K^{\beta\alpha}_{ik}D_l^{\beta\alpha} D_j^{\beta\alpha}+ K^{\beta\alpha}_{jl}D_k^{\beta\alpha} D_i^{\beta\alpha}+ K^{\beta\alpha}_{il}D_k^{\beta\alpha} D_j^{\beta\alpha}. 
\label{eqC2}
\end{equation}
By exluding the rotational degrees of freedom from the analysis, the tensors $M_{ijm}$, $P_{ijmpq}$, $Q_{ijkl}$ and $S_{ijmnl}$ are then set to zero, the couple stress tensor vanishes and the stress tensor reduces to
\begin{equation}
\sigma_{ij}=C_{ijkl}\epsilon_{kl}+E_{ijklmn}\frac{\partial^2 \epsilon_{kl}}{\partial x_m\partial x_n}+F_{ijklmn}\frac{\partial^2 u_{[k,l]}}{\partial x_m\partial x_n}, 
\end{equation}
which is similar to the second-order gradient model \cite{chang1995IJSS, suiker2001-1}. \\
By keeping only the first-order gradient in the analysis, which restricts the order of derivatives of $\textbf{u}$ and $\textbf{w}$ to $\{2,1\}$, the tensors $E_{ijklmn}$, $F_{ijklmn}$ and $P_{ijmpq}$ are then set to zero. The stress tensor and the couple stress tensor then reduce to
\begin{eqnarray}
\sigma_{ij} &= & C_{ijkl}\epsilon_{kl}+M_{ijm}\Pi_m, \label{eqstresscoss} \\
\mu_{ij} & = & Q_{ijkl} \frac{\partial \Pi_k}{\partial x_l} + S_{ijnml} \frac{\partial u_{[m,n]}}{\partial x_l}, 
\label{eqcouplestresscoss}
\end{eqnarray}
which are the constitutive relations of the Cosserat model. \\
An important point is lying here. Considering the frictional case, where $G_B=G_T=0$, the couple stress tensor is then excluded from the analysis, i.e., $\mu_{ij}=0$. The stress tensor in Eq. (\ref{eqstresscoss}) can be found using the following expansion of the discrete displacements and rotations 
\begin{eqnarray}
u_i^{\beta} = u_i^{\alpha}+D_j^{\beta\alpha}\frac{\partial u_i^{\alpha}}{\partial x_j}, \:
w_i^{\beta} = w_i^{\alpha}, 
\label{eqCossexpansion}
\end{eqnarray}
which restricts the order of derivatives of $\textbf{u}$ and $\textbf{w}$ to $\{1,0\}$. 
Thus, the Cosserat model does not contain a derivative over space coordinates in the rotational contribution to translational shear interaction in the stress tensor, which is represented by the term $M_{ijm} \Pi_m$ in Eq. (\ref{eqstresscoss}) and involves the shear stiffness $K_S$. In others words, a dependence on space coordinates of the particle rotations is not taken into account in the prediction of the stress tensor. 
Although this activates the sliding resistance at the surface of the particles as it can be seen in Fig. \ref{figinteraction}(a), it can be expected that the absence of a derivative over space coordinates can lead to an inconsistent representation of the interaction due to the activation of the sliding resistance by particle rotations, and thus, to inconsistent results as it will be further discussed in the next section. On the opposite side, it is important to note that the couple stress tensor $\mu_{ij}$ for the interactions due to rolling resistance, i.e., involving $G_B$ , and due to twisting resistance, i.e., involving $G_T$ contains a derivative over the space coordinates and remains the same in the Cosserat and in the second-order gradient micropolar model. In the next section, a Cosserat model is used to predict the bulk mode propagation in the FCC structure and is compared to the long wavelength approximations of the discrete model. 

\section{Dispersion relations of the bulk mode propagating in the FCC structure using the Cosserat model: limits of the model}
\label{secCosserat}

The dispersion relations of the bulk mode propagating in the FCC structure are predicted using the stress tensor in Eq. (\ref{eqstresscoss}) and the couple stress tensor in Eq. (\ref{eqcouplestresscoss}), which correspond to the Cosserat model. The wave propagation is described by the equation of motion in translation 
\begin{equation}
\rho\frac{\partial^2 u_i}{\partial t^2} = \frac{\partial \sigma_{ji}}{\partial x_j} = C_{ijkl}\frac{\partial \epsilon_{kl}}{\partial x_j}+M_{jim}\frac{\partial \Pi_m}{\partial x_j}, 
\label{eqmottransCoss}
\end{equation}
where $\rho=\eta \rho_b$ and $\eta$ is the volume fraction of the structure, which is $\eta=\pi\sqrt{2}/6\simeq0.74$ in the FCC structure. The equation of rotational motion is
\begin{eqnarray}
J\frac{\partial^2 w_i}{\partial t^2} & = & \frac{\partial \mu_{ji}}{\partial{x_j}}+\varepsilon_{ipq}\sigma_{pq} \nonumber \\
& = & Q_{jikl}\frac{\partial^2 \Pi_k}{\partial x_j\partial x_l }+S_{jinml}\frac{\partial^2 u_{[m,n]} }{\partial x_j \partial x_l}+ \varepsilon_{ipq}(C_{pqrs}\epsilon_{rs}+M_{pqt}\Pi_t). 
\label{eqmotrotCoss}
\end{eqnarray}
where $J=\sqrt{2}I_b/a^3$ is the density of moment of inertia, which is deduced from the number $n$ of beads in a cube with an edge-length $a$ \cite{merkel2011prl}. \\
In order to evaluate $V_{cell}$, The number  $n$ of beads in a cube with an edge-length $a$ is found from the volume fraction $\eta$ and the volume of one bead $V_{bead}$ as
\begin{equation}
n=\eta a^3/V_{bead}=\sqrt{2}\pi a^3/[6(4\pi(a/2)^3/3)]=\sqrt{2}, 
\end{equation} 
i.e., there are $\sqrt{2}$ beads in a cube of volume $a^3$. In others words, the defined volume of the unit cell, which should include 2 beads, is therefore $V_{cell}=a^3\sqrt{2}$. \\
By combination of Eqs. (\ref{eqDab}), (\ref{eqvecposition}) and (\ref{eqC2}), the components of the matrix $C_{IJ}$, which gives the 36 independent constants of the tensor $C_{ijkl}$ \cite{Broyer2000}, in the FCC structure are
\begin{eqnarray}
C_{IJ}=
\left[ \begin{array}{cccccc}
\frac{\sqrt{2}(K_N+K_S)}{a} & \frac{K_N-K_S}{a\sqrt{2}} & \frac{K_N-K_S}{a\sqrt{2}} & 0 & 0 & 0 \\
\frac{K_N-K_S}{a\sqrt{2}} & \frac{5K_N+3K_S}{2a\sqrt{2}} & \frac{K_N-K_S}{2a\sqrt{2}} & 0 & 0 & 0 \\
\frac{K_N-K_S}{a\sqrt{2}} & \frac{K_N-K_S}{2a\sqrt{2}} &  \frac{5K_N+3K_S}{2a\sqrt{2}} & 0 & 0 & 0 \\
0 & 0 & 0 & \frac{K_N+K_S}{a\sqrt{2}} & 0 & 0 \\
0 & 0 & 0 & 0 & \frac{K_N+K_S}{a\sqrt{2}} & 0 \\
0 & 0 & 0 & 0 & 0 & \frac{K_N+3K_S}{2a\sqrt{2}} \\
\end{array}
\right], 
\end{eqnarray}
where $(I,J)\in\{11,22,33,12,13,23\}$. 
By combination of  Eqs.  (\ref{eqDab}), (\ref{eqvecposition}) and (\ref{eqM}), the component of the tensor $M_{ijm}$ reduces to
\begin{equation}
M_{ijm}=-\varepsilon_{ijm}\frac{2\sqrt{2}K_S}{a}. 
\label{eqMtensorFCC}
\end{equation}
The tensors $Q_{jikl}$ and $S_{jinml}$ are more complex and are predicted with the combination of Eqs. (\ref{eqDab}), (\ref{eqvecposition}) with (\ref{eqQtensorFCC}) and (\ref{eqStensorFCC}), respectively, see Sec. \ref{subsecCOHcoss}. 

\subsection{Frictional case}
\label{secDRCoss}

In this section, a frictional assembly is considered. Thus, $G_T=G_B=\mu_{ij}=0$ and the equation of rotational motion reduces to 
\begin{eqnarray}
J\frac{\partial^2 w_i}{\partial t^2} = \varepsilon_{ipq}(C_{pqrs}\epsilon_{rs}+M_{pqt}\Pi_t). 
\label{eqmotrotCossNC}
\end{eqnarray}
Therefore, this case does not contain all the relative degrees of motion of the Cosserat model and can be called a reduced Cosserat model \cite{kulesh2009}.

\subsubsection{Longitudinal wave propagation in direction $x_1=x$}

The axis perpendicular to the cubic layers is the four-fold axis of high symmetry. The calculations here consider only plane wave propagation with the polarization of the wave being pure. Thus, without loss of generality, the calculations do not require the use of scalar $\phi$ and vector $\psi$ potentials such as $\textbf{u}=\nabla\cdot\phi+\nabla\wedge \psi$ \cite{Broyer2000}; it can be checked that the use of these two potentials will lead to the same results.
Writing the equations of motion (\ref{eqmottransCoss}) and (\ref{eqmotrotCossNC}) considering that the only non-zero component of displacement is the component $u_1$ gives
\begin{equation}
\rho \frac{\partial^2 u_1}{\partial t^2}=C_{11}\frac{\partial^2 u_1}{\partial x_1^2}, \quad J \frac{\partial^2 w_i}{\partial t^2}=0. 
\end{equation}
As expected, there is no contribution of the rotational degrees of freedom in the case of a pure longitudinal wave propagation. Considering a plane wave solution $u_1=U_1e^{i(\omega t-k_xx)}$, the dispersion relation for the longitudinal wave is found
\begin{eqnarray}
\omega_L  =  \sqrt{\frac{\sqrt{2}(K_N+K_S)}{a \rho}}k_x = \sqrt{\frac{6(K_N+K_S)}{\pi a \rho_b}}k_x, 
\end{eqnarray}
which is equal to the long wavelength approximation of the discrete model in Eq. (\ref{eqdispdiscL}). In order to compare the prediction from the Cosserat model with the approximations of the discrete model, the continuum mass density $\rho$ is replaced by the mass density of the materials constituting the beads $\rho_b$ and the volume fraction $\eta$. 

\subsubsection{Pure rotational wave propagation in direction $x_1=x$}
\label{secrotcoss}

Considering a pure rotational wave propagating along $x_1$ direction, where $\textbf{u}=0$, Eqs. (\ref{eqmottransCoss}) and (\ref{eqmotrotCoss}) become
\begin{eqnarray}
M_{1im}\frac{\partial \Pi_m}{\partial x_1} & = & 0 \quad \Rightarrow m=1,\nonumber\\
J\frac{\partial^2 w_1}{\partial t^2} & = & \varepsilon_{1jk}M_{jk1}\Pi_1=(M_{231}-M_{321})\Pi_1=-2M_{321}\Pi_1. 
\label{eqpurerot}
\end{eqnarray}
where the index $m$ of the particle spin $\Pi_m$ should be equal in both equations. With $w_i=\Pi_i+\Gamma_i=\Pi_i-\varepsilon_{ijk}u_{[j,k]}/2$ and $u_{[j,k]}=0$, Eq. (\ref{eqpurerot}) then becomes 
\begin{equation}
J\frac{\partial^2 \Pi_1}{\partial t^2}=-\frac{4\sqrt{2}}{a}K_S \Pi_1, 
\end{equation}
which is not an equation describing a wave propagation. 
With the plane wave substitution $\Pi_1=P_1e^{i(\omega t-k_xx)}$, the dispersion relation for the pure rotational mode is found
\begin{equation}
\omega_R=\sqrt{\frac{4\sqrt{2}}{aJ}K_S}=\sqrt{\frac{240K_S}{\pi\rho_ba^3}}, 
\end{equation}
which corresponds to Eq. (\ref{eqdispdiscR}) only for $k_x=0$. This dispersion relation does not depend on $\textbf{k}$, the pure rotational mode does not propagate in the $x_1$ direction, which is a consequence of Eqs. (\ref{eqstresscoss}) and (\ref{eqmotrotCoss}). Since there is no derivative over space coordinates in the rotational contribution to the shear displacement interaction in the stress tensor, there is no propagation of pure rotational wave predicted in the frictional case when only this interaction is taken into account. Thus, the plane wave substitution for the rotational motion in this case is questionable. The case of propagation of a pure rotational wave in a noncohesive Cosserat medium clearly highlights the drawback of the Cosserat model, which is an incomplete representation of the rotational interactions between the particles and, as a consequence, results to an inconsistent prediction of the dispersion relation of bulk mode propagation. 

\subsubsection{Transverse-rotational and rotational-transverse wave propagation in direction $x_1=x$}
\label{sectrrtcoss}

\begin{figure}[ht!]
\centering
\includegraphics[width=8cm]{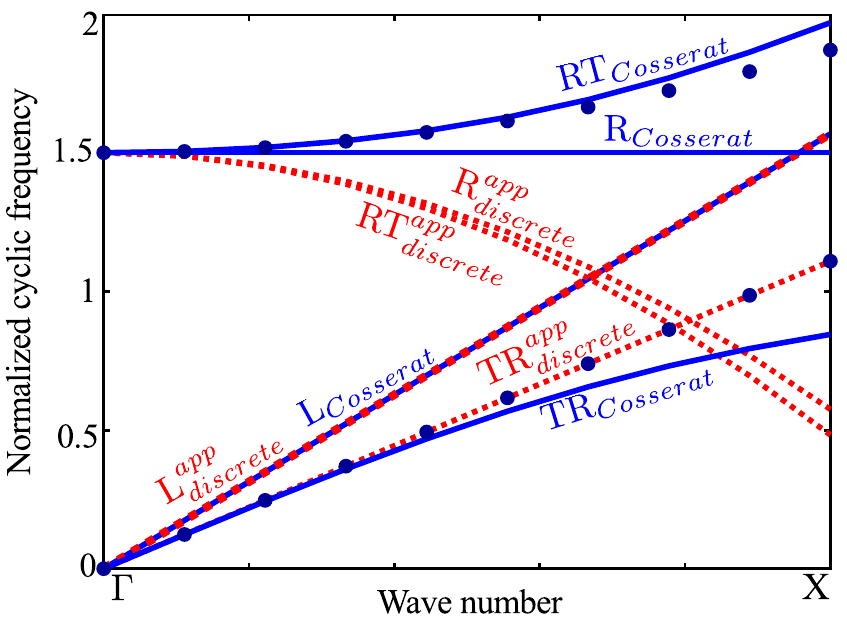}
\caption{(Color online) Dispersion relations in the $x$ direction of the FCC structure from the approximation of the discrete model (red dashed lines), see Sec. \ref{subsecdispdisc}, from the Cosserat model (blue lines), and the approximations for $k_x\rightarrow0$ of the Cosserat model for the modes TR and RT (blue dots). The angular frequencies are normalized with the cutoff frequency of the longitudinal mode $\omega_L^c$ in Eq. (\ref{eqcutoffL}). The parameter used here is $\nu=0.3$, i.e., $\Delta_K\simeq0.82$. }
\label{figCossdispersion}
\end{figure}

The direction $x_1$ is the direction of transverse isotropy in the FCC structure. The case of transverse waves needs the consideration of either $y$ or $z$ components of the displacement. Considering the displacement component $u_y=u_2$, the equation of motion for translation includes the elastic constant $C_{44}$ (which is equal to the elastic constant $C_{55}$ in the case where the displacement component $u_z=u_3$ is considered). From the translational component 2 and the axis of propagation 1, the component of the rotation direction should be 3 and the elements of $M_{ijk}$ involved are $M_{213}$ and $M_{123}$, the equation of motion for translation is then
\begin{eqnarray}
\rho \frac{\partial^2 u_2}{\partial t^2}=C_{44}\frac{\partial}{\partial x_1}(\epsilon_{12}+\epsilon_{21}) - M_{213}\frac{\partial \Pi_3}{\partial x_1}, 
\label{eqmottransshear}
\end{eqnarray}
and the equation of motion for rotation becomes
\begin{eqnarray}
J\frac{\partial^2 w_3}{\partial t^2}=\varepsilon_{312}\sigma_{12}+\varepsilon_{321}\sigma_{12}=-2M_{213}\Pi_3. 
\label{eqmotrotshear}
\end{eqnarray}
In the case of plane wave propagation along the $x_1$ direction, i.e., $\partial u_1/\partial x_2=0$, we can write
\begin{eqnarray}
\epsilon_{12}+\epsilon_{21} & = & \partial{u_2}/\partial x_1, \nonumber \\
w_3 & = & \Pi_3+\Gamma_3=\Pi_3-\varepsilon_{312}u_{[1,2]}/2-\varepsilon_{321}u_{[2,1]}/2 
=  \Pi_3+ \frac{1}{2}\frac{\partial u_2}{\partial x_1} . 
\label{eqintshear}
\end{eqnarray}
Putting Eq. (\ref{eqintshear}) in Eqs. (\ref{eqmottransshear}) and (\ref{eqmotrotshear}), the equations of motion then read
\begin{equation}
\rho\frac{\partial^2 u_2}{\partial t^2}=C_{44}\frac{\partial^2 u_2}{\partial x_1^2}-M_{213}\frac{\partial \Pi_3}{\partial x_1}, 
\label{eqmotS}
\end{equation}
and, 
\begin{equation}
J\biggl(\frac{\partial^2 \Pi_3}{\partial t^2}+ \frac{1}{2}\frac{\partial^2}{\partial t^2}\frac{\partial u_2}{\partial x_1}\biggl)=-2M_{213}\Pi_3. 
\label{eqmotrotS}
\end{equation}
After a plane wave substitution with $u_2=U_2e^{i(wt-k_xx)}$ and $\Pi_3=P_3e^{i(wt-k_xx)}$, the following characteristic equation is found
\begin{equation}
2\rho J \omega^4 - (JM_{213}k_x^2+2JC_{44}k_x^2+4\rho M_{213})\omega^2+4M_{213}C_{44}k_x^2=0. 
\end{equation}
The roots of the characteristic equation give the dispersion relations of the TR and RT modes
\begin{eqnarray}
\omega_{TR,RT}^2 & = & \frac{2C_{44}+M_{213}}{4\rho}k_x^2+\frac{ M_{213}}{J} \pm \biggl[ \biggl(\frac{2C_{44}+M_{213}}{4\rho}k_x^2+\frac{ M_{213}}{J}\biggl)^{2} -\frac{2M_{213}C_{44}}{\rho J}k_x^2\biggl]^{\frac{1}{2}} \nonumber\\
&=&\frac{3(K_N+3K_S)}{2\pi a \rho_b}k_x^2+\frac{120K_S}{\pi \rho_b a^3} \nonumber \\
&  & \pm \biggl[\biggl(\frac{3(K_N+3K_S)}{2\pi a \rho_b} k_x^2+\frac{120K_S}{\pi \rho_b a^3}\biggl)^2 -\frac{720(K_N+K_S)K_S}{\pi^2a^4\rho_b^2}k_x^2 \biggl]^{\frac{1}{2}}, 
\label{eqdispCossTRRT}
\end{eqnarray}
where the minus and plus signs correspond to the TR and RT modes, respectively. 
These dispersion relations corresponds to the ones of \cite{suiker2001-2} when the torque interactions are neglected ($G_B=G_T=0$). A comparison of the dispersion relations from the Cosserat theory with the ones from the discrete theory approximation is exposed in Fig. \ref{figCossdispersion}. 
In the limit $k_x\rightarrow0$, the dispersion relation of the mode TR in Eq. (\ref{eqdispCossTRRT}) becomes
\begin{equation}
\omega_{TR}\simeq\sqrt{\frac{C_{44}}{\rho}}k_x=\sqrt{\frac{3(K_N+K_S)}{\pi a \rho_b}}k_x, 
\end{equation}
which does correspond to the long wavelength approximation of the discrete model in Eq. (\ref{eqdispdiscTR}). This dispersion relation also present a cutoff pulsation in the limit $k_x\rightarrow\infty$
\begin{equation}
\omega_{TR}^c=2\sqrt{\frac{M_{213}C_{44}}{J(M_{213}+2C_{44})}}=\frac{4}{a}\sqrt{\frac{240(K_N+K_S)K_S}{\pi \rho_b a (K_N+3K_S)}}, 
\end{equation}
which does not correspond to the one predicted by the discrete model. The cutoff frequency predicted by the Cosserat theory does not result from the discrete nature of the material since only the first-order gradient is considered. Here, the cutoff frequency is due to the mode repulsion with the rotational-transverse mode and is even larger than the cutoff frequency of the longitudinal mode predicted by the discrete model. Therefore, the cutoff frequency predicted by the Cosserat theory is not physically meaningful. 
For $k_x=0$ the angular pulsation of the rotational-transverse mode is 
\begin{equation}
\omega_{RT}(0)=\sqrt{\frac{2M_{213}}{J}}=\sqrt{\frac{240K_S}{\pi \rho_b a^3}}, 
\end{equation}
which corresponds to the one in the discrete model. In the limit $k_x\rightarrow0$, the dispersion relation of the mode RT in Eq. (\ref{eqdispCossTRRT}) becomes 
\begin{equation}
\omega_{RT} \simeq \sqrt{\frac{2M_{213}}{J}}\biggl(1+\frac{J}{8\rho}k_x^2 \biggl) = \sqrt{\frac{240K_S}{\pi\rho_ba^3}}\biggl( 1 + \frac{a^2}{80}k_x^2\biggl), 
\end{equation}
which does not correspond to the long wavelength approximation of the discrete model in Eq. (\ref{eqdispdiscRT}). 
 In the limit $k_x\rightarrow\infty$, the dispersion relation of the mode RT in Eq. (\ref{eqdispCossTRRT}) becomes 
\begin{equation}
\omega_{RT} \simeq \sqrt{\frac{M_{213}+2C_{44}}{2\rho}}k_x=\sqrt{\frac{3(K_N+3K_S)}{\pi a \rho_b}}k_x, 
\end{equation}
which has no equivalent in the discrete model. Finally, in the Cosserat theory, the dispersion relation of the mode RT is concave upward while it is concave downward in the discrete model as already noticed in \cite{merkel2011prl}. The slope of the mode RT is due to mode repulsion with the mode TR and is not physically meaningful. 
As a consequence, the cutoff frequencies of wave propagation are not correctly predicted by the Cosserat model as well as the phase and group velocities as it can be seen in Fig. \ref{figCossdispersion}. 

In conclusion, the improvement of the Cosserat theory from the classical theory of elasticity is to predict the existence of the rotational and rotational-transverse modes. But the Cosserat model is only able to predict the angular frequency of these modes for $\textbf{k}=0$. The Cosserat model is unable to predict correctly the dispersion relations of these two new modes because the representation of the interaction due to the activation of the sliding resistance by the particle rotations is inconsistent. The case of a rolling and twisting resistant assembly is now considered to show that the Cosserat model suffers from the same drawback in this case. 

\subsection{Rolling and Twisting resistant case}
\label{subsecCOHcoss}

The bulk mode propagation along the axis $x=x_1$ is considered following the derivations in Sec. \ref{secDRCoss}. The dispersion of the longitudinal wave remains unchanged since there is no rotational motion involved for this mode. 

\subsubsection{Pure rotational wave in direction $x_1=x$}
\label{secrotcossCOH}

Considering a pure rotational wave propagating along $x_1$ direction, the equation of motion for rotation is
\begin{eqnarray}
J\frac{\partial^2 w_1}{\partial t^2} & = & Q_{1111}\frac{\partial^2 \Pi_1}{\partial x_1^2}+ \varepsilon_{1jk}M_{jk1}\Pi_1=Q_{1111}\frac{\partial^2 \Pi_1}{\partial x_1^2}-2M_{321}\Pi_1. 
\label{eqpurerotCOH}
\end{eqnarray}
With $Q_{1111}=\sqrt{2}(G_B+G_T)/a$, $w_i=\Pi_i+\Gamma_i=\Pi_i-\varepsilon_{ijk}u_{[j,k]}/2$ and $u_{[j,k]}=0$, Eq. (\ref{eqpurerotCOH}) then read
\begin{equation}
J\frac{\partial^2 \Pi_1}{\partial t^2}=-\frac{4\sqrt{2}}{a}K_S \Pi_1+\frac{\sqrt{2}}{a}(G_B+G_T)\frac{\partial^2 \Pi_1}{\partial x_1^2}.  
\end{equation}
With the plane wave substitution $\Pi_1=P_1e^{i(\omega t-k_xx)}$, the dispersion relation for the pure rotational mode is found
\begin{equation}
\omega_R^2=\frac{4\sqrt{2}K_S}{aJ}+\frac{\sqrt{2}(G_B+G_T)}{aJ}k_x^2=\frac{240K_S}{\pi\rho_ba^3}+\frac{60(G_B+G_T)}{\pi\rho_ba^3}k_x^2. 
\end{equation}
The Taylor expansion for $k_x\rightarrow0$ is
\begin{equation}
\omega_R=\sqrt{\frac{240K_S}{\pi\rho_ba^3}}\biggl(1+\frac{G_B+G_T}{8K_S}k_x^2\biggl)
\label{eqRcossCOH}
\end{equation}
which corresponds to the long wavelength approximation of the discrete model in Eq. (\ref{eqdispdiscRCOH}) only when $k_x=0$. Comparing Eqs. (\ref{eqdispdiscRCOH}) and (\ref{eqRcossCOH}), the term depending on the stiffnesses $G_B$ and $G_T$ are equal. In contrast, the term depending only on the stiffness $K_S$ and the wavenumber $k_x$ differs. Thus, the prediction from the Cosserat model does not correspond to the approximate dispersion relation of the discrete model in Eq. (\ref{eqdispdiscRCOH}). 

\subsubsection{Transverse-rotational and rotational-transverse wave in direction $x_1=x$}
\label{sectrrtcossCOH}

For the case of the wave involving the displacement $u_2$ and the rotation $\Pi_3$, the equation of motion for translation is
\begin{eqnarray}
\rho \frac{\partial^2 u_2}{\partial t^2}=C_{44}\frac{\partial}{\partial x_1}(\epsilon_{21}+\epsilon_{12}) - M_{213}\frac{\partial \Pi_3}{\partial x_1}. 
\label{eqmottransshearCOH}
\end{eqnarray}
The equation of motion for rotation is
\begin{eqnarray}
J\frac{\partial^2 w_3}{\partial t^2} & = & Q_{1331}\frac{\partial^2 \Pi_3}{\partial x_1^2}+S_{13121}\frac{\partial^2 u_{[2,1]}}{\partial x_1^2}+\varepsilon_{3jk}\sigma_{jk} \nonumber \\
& = &-2M_{213}\Pi_3 + Q_{1331}\frac{\partial^2 \Pi_3}{\partial x_1^2}+S_{13121}\frac{\partial^2 u_{[2,1]}}{\partial x_1^2}, \nonumber \\
\label{eqmotrotshearCOH}
\end{eqnarray}
with $Q_{1331}=(G_T+3G_B)/(a\sqrt{2})$ and $S_{13121}=(G_T+3G_B)/(a\sqrt{2})$. 
Putting Eq. (\ref{eqintshear}) in Eqs. (\ref{eqmottransshearCOH}) and (\ref{eqmotrotshearCOH}), the equations of motion then read
\begin{equation}
\rho\frac{\partial^2 u_2}{\partial t^2}=C_{44}\frac{\partial^2 u_2}{\partial x_1^2}-M_{213}\frac{\partial \Pi_3}{\partial x_1}, 
\label{eqmotSCOH}
\end{equation}
and 
\begin{equation}
J\biggl(\frac{\partial^2 \Pi_3}{\partial t^2}+ \frac{1}{2}\frac{\partial^2}{\partial t^2}\frac{\partial u_2}{\partial x_1}\biggl)=-2M_{213}\Pi_3+ Q_{1331}\frac{\partial^2 \Pi_3}{\partial x_1^2}+\frac{1}{2}S_{13121}\frac{\partial^3 u_2}{\partial x_1^3}. 
\label{eqmotrotSCOH}
\end{equation}
After a plane wave substitution with $u_2=U_2e^{i(wt-k_xx)}$ and $\Pi_3=P_3e^{i(wt-k_xx)}$, the following characteristic equation is found
\begin{eqnarray}
2\rho J \omega^4 &-& (JM_{213}k_x^2+2JC_{44}k_x^2+4\rho M_{213}+2\rho Q_{1331}k^2_x)\omega^2 \nonumber \\
&+ &4M_{213}C_{44}k_x^2+2C_{44}Q_{1331}k_x^4+M_{213}S_{13121}k_x^4=0. 
\end{eqnarray}
The roots of the characteristic equation give the dispersion relations of the TR and RT modes 
\begin{eqnarray}
\omega_{TR,RT}^2 & = & \biggl(\frac{2C_{44}+M_{213}}{4\rho}+\frac{Q_{1331}}{2J}\biggl)k_x^2+\frac{ M_{213}}{J}
\pm \biggl\{ \biggl[\biggl(\frac{2C_{44}+M_{213}}{4\rho}+\frac{Q_{1331}}{2J}\biggl)k_x^2 \nonumber \\
& & +\frac{ M_{213}}{J}\biggl]^{2} -\frac{2M_{213}C_{44}}{\rho J}k_x^2-\frac{2C_{44}Q_{1331}+M_{213}S_{13123}}{2\rho J}k_x^4\biggl\}^{\frac{1}{2}} \nonumber\\
&=&\frac{3(K_N+3K_S)}{2\pi a \rho_b}k_x^2+\frac{15(3G_B+G_T)}{\pi \rho_ba^3}k_x^2+\frac{120K_S}{\pi \rho_b a^3} \nonumber\\
& &\pm\biggl[\biggl(\frac{3(K_N+3K_S)}{2\pi a \rho_b} k_x^2+\frac{15(3G_B+G_T)}{\pi \rho_ba^3}k_x^2+\frac{120K_S}{\pi \rho_b a^3}\biggl)^2 \nonumber \\
& & -\frac{720(K_N+K_S)K_S}{\pi^2a^4\rho_b^2}k_x^2 - 90(3G_B+G_T)\frac{K_N+3K_S}{\pi^2a^4\rho_b^2}k_x^4\biggl]^{\frac{1}{2}}, 
\label{eqdispCossTRRTCOH} 
\end{eqnarray}
where the minus and plus signs correspond to the TR and RT modes, respectively. 
In the limit $k_x\rightarrow0$, the dispersion relation (\ref{eqdispCossTRRTCOH}) for the mode TR becomes
\begin{equation}
\omega_{TR}\simeq\sqrt{\frac{C_{44}}{\rho}}k_x=\sqrt{\frac{3(K_N+K_S)}{\pi a \rho_b}}k_x, 
\end{equation}
which is equal to the long wavelength approximation of the discrete model in Eq. (\ref{eqdispdiscTRCOH}). 
For $k_x=0$ the angular frequency of the rotational-transverse mode is 
\begin{equation}
\omega_{RT}(0)=\sqrt{\frac{2M_{213}}{J}}=\sqrt{\frac{240K_S}{\pi \rho_b a^3}}, 
\end{equation}
which corresponds to the one of the discrete theory. In the limit $k_x\rightarrow0$, the dispersion relation (\ref{eqdispCossTRRTCOH}) for the mode RT becomes 
\begin{equation}
\omega_{RT}\simeq\sqrt{\frac{2M_{213}}{J}}\biggl[1+\biggl(\frac{J}{8\rho}+\frac{Q_{1331}}{4M_{213}} \biggl)k_x^2\biggl]=\sqrt{\frac{240K_S}{\pi a^3 \rho_b}}\biggl[1+\biggl(\frac{a^2}{80}+\frac{3G_B+G_T}{16K_S}\biggl)k_x^2 \biggl], 
\label{eqRTcossCOH}
\end{equation}
where, once again, the term depending on the stiffnesses $G_B$ and $G_T$ correspond to the one of the approximation of the discrete mode in Eq. (\ref{eqdispdiscRTCOH}), whereas the terms depending only on $K_S$ and $k_x$ differ. 
\begin{figure}[ht!]
\centering
\includegraphics[width=14cm]{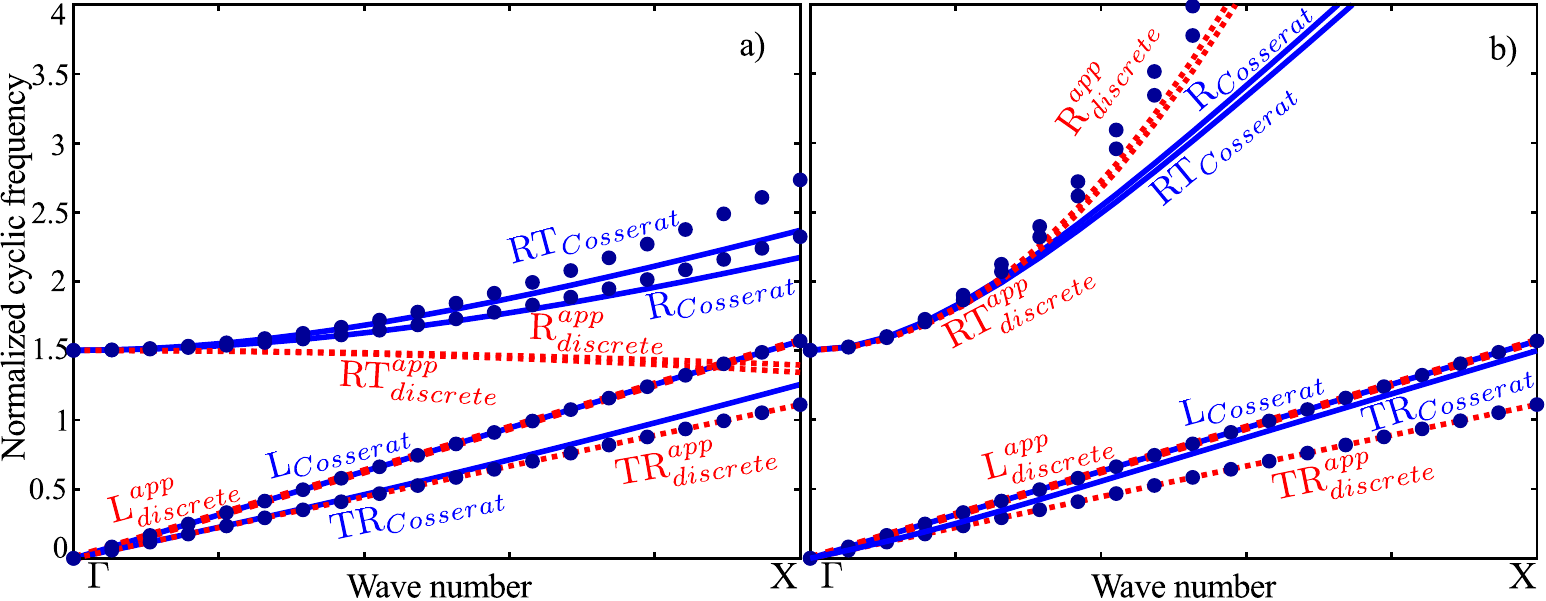}
\caption{(Color online) Dispersion relations in the $x_1$ direction of the rolling and twisting resistant FCC structure from the approximation of the discrete model (red dashed lines), see Sec. \ref{subsecdispdiscCOH}, the Cosserat model (blue continuous lines) and the Cosserat model approximations (blue dots). The parameters here are $\Delta K=K_S/K_N \simeq 0.82$ and $a=2$~mm. (a) $G_B=K_Na^2/10$ and $G_T=K_Sa^2/10$. (b) $G_B=K_Na^2$ and $G_T=K_Sa^2$. The cyclic frequencies are normalized with the cutoff frequency of the longitudinal mode $\omega_L^c$ in Eq. (\ref{eqcutoffL}). }
\label{figcosseratdispersionCOH}
\end{figure}

In conclusion to the analysis in the rolling and twisting resistant case, the torque interactions, involving the bending stiffness $G_B$ and the twisting stiffness $G_T$, are correctly represented by the Cosserat model in comparison of the approximations of the discrete model. However, as it can be seen in Fig. \ref{figcosseratdispersionCOH}, the dispersion curves predicted by the Cosserat model still do not correspond to the ones of the discrete model approximation. 
The Cosserat model predicts correctly only the cutoff frequencies of the rotational-related modes (R and RT) when the wavenumber is $\textbf{k}=0$. The phase and group velocities are therefore incorrectly predicted by the Cosserat model. The drawback of the Cosserat modeling concerns therefore only the shear interaction between the beads, and more specifically the contribution of the particle rotations to it. \\

To conclude the analysis of the Cosserat model, its drawback is identified. The interaction between the particles related to the rotational contribution to the stress tensor is not correctly represented. This comes from the absence of a derivatives over space coordinates in the term involving the particle rotation in the stress tensor, i.e., the term involving $\Pi_i$ in Eq. (\ref{eqstresscoss}). \\
This drawback can be overcome using the stress tensor resulting from the second-order gradient derivation, as done below in the proposed enhanced micropolar model. 


\section{Dispersion relations using the enhanced micropolar model}
\label{secEnhanced}

In the stress tensor in Eq. (\ref{eqSGstress}), which is derived from the second-order gradient micropolar model, the rotational contribution to the translational shear interaction between the particles is included in the terms depending on $M_{ijm} $ and $P_{ijmpq}$. For this interaction, derivatives over space coordinates are introduced through the term relative to the elastic tensor $P_{ijmpq}$. Nevertheless, the second-order gradient model introduces two other sixth order elastic tensors, namely $E_{ijklmn}$ and $F_{ijklmn}$. As a consequence, a continuum, even if isotropic, is described by a large number elastic constants, which induces a complexity in the modeling that could be considered as disqualifying. Here, it is assumed that the influence of the terms related to the tensors $E_{ijklmn}$ and $F_{ijklmn}$ is negligible at least for small wavenumbers (or large wavelengths). These two tensors are thus set to zero in order to reduce the number of elastic constants involved in the stress tensor, i.e.,  the stress tensor of the enhanced micropolar model is 
\begin{equation}
\sigma_{ij}=C_{ijkl}\epsilon_{kl}+M_{ijm}\Pi_m+P_{ijmpq}\frac{\partial^2 \Pi_m}{\partial x_p\partial x_q}, 
\label{eqEMstress}
\end{equation}
with a new elastic tensor, relative to Eq. (\ref{eqstresscoss}), $P_{ijmpq}$, which is defined in Eq. (\ref{eqP}). 
The couple stress tensor remains equal to the one in Eq. (\ref{eqSGcouplestress}), which is the same as the couple stress tensor of the Cosserat model in Eq. (\ref{eqcouplestresscoss}), which is
\begin{equation}
\mu_{ij}=Q_{ijkl}\frac{\partial \Pi_k}{\partial x_l}+S_{ijnml}\frac{\partial u_{[m,n]}}{\partial x_l}
\end{equation}
Thus, the equation of motion in translation of the enhanced micropolar model is
\begin{equation}
\rho\frac{\partial^2 u_i}{\partial t^2} = \frac{\partial \sigma_{ji}}{\partial x_j} = C_{ijkl}\frac{\partial \epsilon_{kl}}{\partial x_j}+M_{jim}\frac{\partial \Pi_m}{\partial x_j}+P_{jimpq}\frac{\partial^3 \Pi_m}{\partial x_p\partial x_q\partial x_j},   
\label{eqmottransCossCOHEM}
\end{equation}
and the equation of motion in rotation is 
\begin{eqnarray}
J\frac{\partial^2 w_i}{\partial t^2} & = & \frac{\partial m_{ji}}{\partial{x_j}}+\varepsilon_{ipq}\sigma_{pq} 
 =  Q_{jikl}\frac{\partial^2 \Pi_k}{\partial x_j\partial x_l }+S_{jinml}\frac{\partial^2 u_{[m,n]} }{\partial x_j \partial x_l}\nonumber \\
 & & + \varepsilon_{ipq}(C_{pqrs}\epsilon_{rs}+M_{pqt}\Pi_t+P_{pqtuv}\frac{\partial^2 \Pi_t}{\partial x_u\partial x_v}). 
\label{eqmotrotCossCOHEM}
\end{eqnarray}

The pure longitudinal wave dispersion relation will not be changed with this model since the rotational motion has no influence on the propagation of this mode. Only the dispersion of the rotational, rotational-transverse and transverse-rotational modes will be considered in the frictional, rolling and twisting resistant cases. 

\subsection{Frictional case}

\subsubsection{Pure rotational wave propagation}

Corresponding to Sec. \ref{secrotcoss} and the expression of the stress tensor in Eq. (\ref{eqEMstress}), the equation of motion for rotation becomes
\begin{eqnarray}
J\frac{\partial^2 \Pi_1}{\partial t^2} & =  &\varepsilon_{1jk}(M_{jk1}\Pi_1+P_{jk111}\frac{\partial^2 \Pi_1}{\partial x_1^2}), \nonumber \\
& =&  -2M_{321}\Pi_1+(P_{23111}-P_{32111})\frac{\partial^2 \Pi_1}{\partial x_1^2}, 
\end{eqnarray}
which contains the derivatives over space coordinates of a wave equation. After a plane wave substitution $\Pi_1=P_1e^{i(\omega t - k_x x)}$, the dispersion relation of the pure rotational wave is
\begin{equation}
\omega^2_R = (2M_{321}+(P_{23111}-P_{32111})k_x^2)/J. 
\end{equation}
After a Taylor expansion in the limit $k_x\rightarrow0$, the dispersion relation becomes
\begin{equation}
\omega_R \simeq \sqrt{\frac{2M_{321}}{J}} \biggl(1+\frac{P_{23111}-P_{32111}}{4M_{321}}k_x^2\biggl). 
\label{eqappRSG1}
\end{equation}
From Eq. (\ref{eqP}), $P_{ij111}=\varepsilon_{ji1}K_S a/(4\sqrt{2})$, Eq. (\ref{eqappRSG1}) is then
\begin{equation}
\omega_R \simeq \sqrt{\frac{240 K_S}{\pi \rho_b a^3}}\biggl(1-\frac{1}{32} k_x^2a^2\biggl), 
\label{eqLWenhR}
\end{equation}
which is exactly the long wavelength approximation of the discrete theory in Eq. (\ref{eqdispdiscR}). 

\subsubsection{Transverse-rotational and rotational-transverse modes}

Corresponding to Sec. \ref{sectrrtcoss} and adding the new component of the stress tensor, the equations of motion (\ref{eqmotS}) and (\ref{eqmotrotS}) then read
\begin{equation}
\rho\frac{\partial^2 u_2}{\partial t^2}=C_{44}\frac{\partial^2 u_2}{\partial x_1^2}-M_{213}\frac{\partial \Pi_3}{\partial x_1} + P_{12311} \frac{\partial^3 \Pi_3}{\partial x_1^3}, 
\label{eqmotSenh}
\end{equation}
and, 
\begin{equation}
J\biggl(\frac{\partial^2 \Pi_3}{\partial t^2}+ \frac{1}{2}\frac{\partial^2}{\partial t^2}\frac{\partial u_2}{\partial x_1}\biggl)=-2M_{213}\Pi_3+(P_{12311}-P_{21311})\frac{\partial^2 \Pi_3}{\partial x_1^2}. 
\label{eqmotrotSenh}
\end{equation}
After a plane wave substitution with $u_2=U_2e^{i(wt-k_xx)}$ and $\Pi_3=P_3e^{i(wt-k_xx)}$, the following characteristic equation is found
\begin{eqnarray}
& & 2\rho J \omega^4-[4\rho M_{213}+JM_{213}k_x^2+2JC_{44}k_x^2-2\rho(P_{21311}-P_{12311})k_x^2+JP_{12311}k_x^4]\omega^2 \nonumber \\
& &+ 4M_{213}C_{44}k_x^2+2C_{44}(P_{12311}-P_{21311})k_x^4=0. 
\end{eqnarray}
The roots of the characteristic equation give the dispersion relation of the TR and RT mode
\begin{eqnarray}
\omega_{TR,RT}^2 & = & \frac{M_{213}}{J}+\biggl(\frac{M_{213}+2C_{44}}{4\rho}-\frac{P_{21311}-P_{12311}}{2J}\biggl)k_x^2+\frac{P_{12311}}{4\rho}k_x^4 \nonumber \\
& \pm & \biggl\{\biggl[\frac{M_{213}}{J}+\biggl(\frac{M_{213}k^2+2C_{44}}{4\rho}-\frac{P_{21311}-P_{12311}}{2J}\biggl)k_x^2+\frac{P_{12311}}{4\rho}k_x^4\biggl]^2 \nonumber \\
& - & \frac{2 M_{213}k_x^2+(P_{12311}-P_{21311})k_x^4}{\rho J}C_{44}\bigg\}^{\frac{1}{2}},  
\end{eqnarray}
where the minus and plus signs corresponds to the TR and RT modes, respectively. 
From Eq. (\ref{eqP}), $P_{12311}=-\sqrt{2}K_Sa/4$ and $P_{21311}=\sqrt{2}K_Sa/8$, the dispersion relations become
\begin{eqnarray}
\omega_{TR,RT}^2 & = &\frac{120 K_S}{\pi \rho_b a^3} +\frac{(6K_N-27K_S)k_x^2}{4\pi \rho_b a}-\frac{3K_Sk_x^4a}{8\pi \rho_b}
\pm \biggl[ \biggl( \frac{120 K_S}{\pi \rho_b a^3} +\frac{(6K_N-27K_S)k_x^2}{4\pi \rho_b a}\nonumber \\
& & -\frac{3K_Sk_x^4a}{8\pi \rho_b}\biggl)^2 
 +  \frac{(K_N+K_S)K_S}{\pi^2a^2\rho_b^2}\biggl(\frac{145 k_x^4}{2}-\frac{720k_x^2}{a^2}\biggl) \biggl]^{\frac{1}{2}}. 
\label{eqdispTRRTenh}
\end{eqnarray}
In the limit $k_x\rightarrow0$, the dispersion relations (\ref{eqdispTRRTenh}) become for the mode TR
\begin{equation}
\omega_{TR}\simeq\sqrt{\frac{C_{44}}{\rho}}k_x=\sqrt{\frac{3(K_N+K_S)}{\pi a \rho_b}}k_x, 
\label{eqLWenhTR}
\end{equation}
and for the mode RT
\begin{eqnarray}
\omega_{RT} &\simeq  &\sqrt{\frac{2M_{213}}{J}}\biggl[1+\biggl(\frac{J}{8\rho}+\frac{P_{12311}-P_{21311}}{4M_{213}}\biggl)k_x^2 \biggl] \nonumber \\
&=& \sqrt{\frac{240 K_S}{\pi \rho_b a^3}}\biggl(1-\frac{11}{320}k_x^2a^2\biggl), 
\label{eqLWenhRT}
\end{eqnarray}
which are exactly the long wavelength approximation of the discrete theory in Eqs. (\ref{eqdispdiscTR}) and (\ref{eqdispdiscRT}), respectively.

\subsection{Rolling and twisting resistant case}

\subsubsection{Pure rotational wave propagation}

Considering Sec. \ref{secrotcossCOH} and applying the enhanced micropolar model,  the equation of motion for rotation becomes
\begin{equation}
J\frac{\partial^2 \Pi_1}{\partial t^2} = Q_{1111}\frac{\partial^2 \Pi_1}{\partial x_1^2}-2M_{321}\Pi_1+(P_{23111}-P_{32111})\frac{\partial^2 \Pi_1}{\partial x_1^2}.   
\end{equation}
The corresponding  dispersion relation is 
\begin{eqnarray}
\omega_R^2 &=  &\frac{2M_{321}}{J}+\frac{P_{23111}-P_{32111}+Q_{1111}}{J}k_x^2 \nonumber \\
&= &\frac{240K_S}{\pi\rho_b a}-\frac{15K_S}{\pi\rho_b a}k_x^2+60\frac{G_B+G_T}{\pi \rho_b a^3}k_x^2. 
\end{eqnarray}
From a Taylor expansion for $k_x\rightarrow0$, it becomes
\begin{eqnarray}
\omega_R &\simeq& \sqrt{\frac{2M_{321}}{J}}\biggl(1+\frac{P_{23111}-P_{23111}+Q_{1111}}{4M_{321}}k^2_x\biggl) \nonumber \\
&= &\sqrt{\frac{240K_S}{\pi\rho_Ba^3}}\biggl(1-\frac{a^2}{32}k_x^2+\frac{G_B+G_T}{8K_S}k_x^2\biggl), 
\label{eqLWenhRCOH}
\end{eqnarray}
which is exactly the long wavelength approximation of the discrete model in Eq. (\ref{eqdispdiscRCOH}). 

\subsection{Transverse-rotational and rotational-transverse modes}

Considering Sec. \ref{sectrrtcossCOH} and applying the enhanced micropolar model, the equation of motion for translation is 
\begin{equation}
\rho \frac{\partial^2 u_2}{\partial t^2}=C_{44}\frac{\partial^2 u_2}{\partial x_1^2}-M_{213}\frac{\partial \Pi_3}{\partial x_1}+P_{12311}\frac{\partial^3 \Pi_3}{\partial x_1^3}. 
\end{equation}
The equation of motion for rotation is 
\begin{equation}
J\biggl(\frac{\partial^2 \Pi_3}{\partial t^2}+ \frac{1}{2}\frac{\partial^2}{\partial t^2}\frac{\partial u_2}{\partial x_1}\biggl)=-2M_{213}\Pi_2+ Q_{1331}\frac{\partial^2 \Pi_3}{\partial x_1^2}+\frac{1}{2}S_{13121}\frac{\partial^3 u_2}{\partial x_1^3}+(P_{12311}-P_{21311})\frac{\partial^2 \Pi_3}{\partial x_1^2}.  
\end{equation}
The corresponding characteristic equation is 
\begin{eqnarray}
 2\rho J \omega^4& -& [JM_{213}k_x^2+2JC_{44}k_x^2+4\rho M_{213}+2\rho Q_{1331}k^2_x-2\rho(P_{21311}-P_{12311})k_x^2+JP_{12311}k_x^4]\omega^2 \nonumber \\
& &+ 4M_{213}C_{44}k_x^2+2C_{44}Q_{1331}k_x^4+M_{213}S_{13121}k_x^4+2C_{44}(P_{12311}-P_{21311})k_x^4=0. 
\end{eqnarray}
The roots of the characteristic equation give the dispersion relations of the TR and RT modes
\begin{eqnarray}
\omega_{TR,RT}^2 & = & \biggl(\frac{2C_{44}+M_{213}}{4\rho}+\frac{Q_{1331}}{2J}-\frac{P_{21311}-P_{12311}}{2J}\biggl)k_x^2+\frac{P_{12311}}{4\rho}k_x^4+\frac{ M_{213}}{J}\nonumber \\
&\pm& \biggl\{ \biggl[\biggl(\frac{2C_{44}+M_{213}}{4\rho}+\frac{Q_{1331}}{2J}-\frac{P_{21311}-P_{12311}}{2J}\biggl)k_x^2+\frac{P_{12311}}{4\rho}k_x^4+\frac{ M_{213}}{J}\biggl]^{2} \nonumber \\
&-& \frac{2M_{213}C_{44}}{\rho J}k_x^2+\biggl[\frac{2C_{44}(P_{12311}-P_{21311}-Q_{1331})+M_{213}S_{13121}}{2\rho J}\biggl]k_x^4\biggl\}^{\frac{1}{2}} \nonumber\\
&=&\frac{3(2K_N-9K_S)}{4\pi a \rho_b}k_x^2+\frac{15(3G_B+G_T)}{\pi \rho_ba^3}k_x^2-\frac{3K_Sa}{8\pi\rho_b}k_x^4+\frac{120K_S}{\pi \rho_b a^3} \nonumber\\
& \pm&\biggl[\biggl(\frac{3(2K_N-9K_S)}{4\pi a \rho_b}k_x^2+\frac{15(3G_B+G_T)}{\pi \rho_ba^3}k_x^2-\frac{3K_Sa}{8\pi\rho_b}k_x^4+\frac{120K_S}{\pi \rho_b a^3}\biggl)^2 \nonumber \\
&-&\frac{(K_N+K_S)K_S}{\pi^2a^4\rho_b^2}\biggl(\frac{145k_x^4}{2}-\frac{720k_x^2}{a^2}\biggl) - 90(3G_B+G_T)\frac{K_N+3K_S}{\pi^2a^4\rho_b^2}k_x^4\biggl]^{\frac{1}{2}}, \nonumber \\
\label{eqdispTRRTenhCOH}
\end{eqnarray}
where the minus and plus signs corresponds to the TR and RT modes, respectively. 
In the limit of small wavenumber $k_x\rightarrow0$, the dispersion relation (\ref{eqdispTRRTenhCOH}) for the mode TR becomes
\begin{equation}
\omega_{TR}\simeq\frac{C_{44}}{\rho}k_x=\sqrt{\frac{3(K_N+K_S)}{\pi a \rho_b}}k_x,  
\label{eqLWenhTRCOH}
\end{equation}
which is exactly the long wavelength approximation in Eq. (\ref{eqdispdiscTRCOH}). In the limit of small wavenumber $k_x\rightarrow0$, the dispersion relation (\ref{eqdispTRRTenhCOH}) for the mode RT becomes
\begin{eqnarray}
\omega_{RT}& \simeq&\sqrt{\frac{2M_{213}}{J}}\biggl[1+\biggl(\frac{J}{8\rho}+\frac{Q_{1331}+P_{12311}-P_{21311}}{4M_{213}} \biggl)k_x^2\biggl]\nonumber\\
&=&\sqrt{\frac{240K_S}{\pi a^3 \rho_b}}\biggl[1+\biggl(\frac{3G_B+G_T}{16K_S}-\frac{11}{320}a^2\biggl)k_x^2 \biggl], 
\label{eqLWenhRTCOH}
\end{eqnarray}
which is exactly the long wavelength approximation of the discrete model in Eq. (\ref{eqdispdiscRTCOH}). 

The equality between Eqs. (\ref{eqdispdiscR}) and (\ref{eqLWenhR}), (\ref{eqdispdiscTR}) and (\ref{eqLWenhTR}), (\ref{eqdispdiscRT}) and (\ref{eqLWenhRT}), (\ref{eqdispdiscRCOH}) and (\ref{eqLWenhRCOH}), (\ref{eqdispdiscTRCOH}) and (\ref{eqLWenhTRCOH}), (\ref{eqdispdiscRTCOH}) and (\ref{eqLWenhRTCOH}) demonstrates that the proposed enhanced micropolar model is a correct description of the long wavelength propagation in a medium accounting for all the rotational degrees of freedom of each particle. The rotational contribution to shear displacement interaction is here correctly modeled with a minimal number of elastic constants needed for a correct description of all the interactions between the particles. 

\section{Conclusion}

The dispersion relations of the bulk modes propagating in a face-centered cubic assembly of monodisperse beads have been theoretically evaluated using three different models. \\
First, the discrete theory, which takes into account all the translational and rotational degrees of freedom of each individual particle, is derived and its approximations for long wavelength are used as references. The constitutive stress and couple stress tensors are derived from the force-displacement and torque-rotation relations of the discrete model through the gradient expansion in particle displacements and rotations, which transform the discrete degrees of freedom into continuum field variables. From this, the stress and couple stress tensors of the Cosserat model and second-order gradient micropolar model are derived. \\
Second, the drawbacks of the Cosserat model are exposed through a direct comparison with the long wavelength approximation of the discrete model, which indicates that one of the interactions between the particles due to rotations is not correctly described. This is the rotational contribution to the shear displacement, which involves the sliding resistance at the surface of the beads. \\
Third, it is demonstrated that the limit of the Cosserat model can be overcome using only one of the additional elastic tensors of the second-order gradient micropolar model; this defines a new model which is called enhanced micropolar model. The long wavelength approximations of the discrete and enhanced micropolar model are exactly equal. The enhanced micropolar model correctly models a granular assembly involving the minimum number of elastic constants possible. The study of perfect crystalline structure is a necessary step to develop a continuum model since a direct comparison with the discrete model is possible only in this case. \\
For disordered systems, only long wavelength waves propagate, so that the present enhanced micropolar model might suffice to describe this case. This work thus contributes to a better modeling of the elastic behavior of realisitic, disordered granular assemblies. It should be possible to compare the simple model also to experimental data and to simulations using
small polydispersity.

\section*{Acknowledgments}
The authors thank Vanessa Magnanimo for the fruitful discussions. This work has been supported by NWO-STW via the VICI project 10828.

\end{document}